\documentclass[a4paper,fleqn,usenatbib]{mnras}

\usepackage{newtxtext,newtxmath}
\usepackage[T1]{fontenc}
\usepackage{ae,aecompl}
\pdfoutput=1

\usepackage{graphicx}	% Including figure files
\usepackage{amsmath}	% Advanced maths commands
\usepackage{amssymb}	% Extra maths symbols

\newcommand{\sa}{SDSS J085319.63$+$162644.4}

\newcommand{\co}{\textit{Chandra} }
\title[Characterisation of a candidate dual AGN]{Characterisation of a candidate dual AGN}

\author[D. Lena et al.]{
D. Lena,$^{1,2}$\thanks{E-mail: d.lena@sron.nl (DL)}
G. Panizo-Espinar,$^{3}$
P. G. Jonker,$^{1,2}$
M. A. P. Torres,$^{4,5,1}$
M. Heida$^{6}$
\\
$^{1}$SRON, Netherlands Institute for Space Research, Sorbonnelaan 2, NL-3584 CA Utrecht, the Netherlands\\
$^{2}$Department of Astrophysics/IMAPP, Radboud University, Nijmegen, PO Box 9010, NL-6500 GL Nijmegen, the Netherlands\\
$^{3}$Facultad de Física, Universidad de La Laguna, Astrofísico Francisco Sánchez s/n, E-38206 La Laguna, Spain\\
$^{4}$Instituto de Astrofisica de Canarias, Calle Via Lactea s/n, E-38205 La Laguna, Tenerife, Spain\\
$^{5}$Dpto. de Astrofísica, Universidad de La Laguna, Astrofísico Francisco Sánchez s/n, E-38206 La Laguna, Spain\\
$^{6}$Cahill Center for Astronomy and Astrophysics, California Institute of Technology, 1200 California Boulevard, Pasadena, CA 91125, USA
}

\date{Accepted for publication in MNRAS.}

\pubyear{2015}

\begin{document}
\label{firstpage}
\pagerange{\pageref{firstpage}--\pageref{lastpage}}
\maketitle

\begin{abstract}
We present \co and optical observations of a candidate dual AGN discovered serendipitously while searching for recoiling black holes via a cross-correlation between the serendipitous XMM source catalog (2XMMi) and SDSS-DR7 galaxies with a separation no larger than ten times the sum of their Petrosian radii. The system has a stellar mass ratio M$_{1}$/M$_{2}\approx 0.7$. One of the galaxies (Source 1) shows clear evidence for AGN activity in the form of hard X-ray emission and optical emission-line diagnostics typical of AGN ionisation. The nucleus of the other galaxy (Source 2) has a soft X-ray spectrum, bluer colours, and optical emission line ratios dominated by stellar photoionisation with a ``composite" signature, which might indicate the presence of a weak AGN. When plotted on a diagram with X-ray luminosity vs [OIII] luminosity both nuclei fall within the locus defined by local Seyfert galaxies. From the optical spectrum we estimate the electron densities finding n$_{1} < 27$ e$^{-}$ cm$^{-3}$ and n$_{2} \approx 200$ e$^{-}$ cm$^{-3}$. From a 2D decomposition of the surface brightness distribution we infer that both galaxies host rotationally supported bulges (Sersic index $< 1$). 
While the active nature of Source 1 can be established with confidence, whether the nucleus of Source 2 is active remains a matter of debate. Evidence that a faint AGN might reside in its nucleus is, however, tantalising.
\end{abstract}


%%%%%%%%%%%%%%%%%%%%%%%%%%%%%%%%%%%%%%%%%%%%%%%%%%

%%%%%%%%%%%%%%%%% BODY OF PAPER %%%%%%%%%%%%%%%%%%
\section{Introduction}
\label{sec: intro}
In the recipe of galaxy-evolution there are two ingredients of key importance: mergers and nuclear activity. Major mergers (defined as having a mass ratio M$_{2}$/M$_{1} \geq 0.3$) affect in dramatic ways the morphologies, star-formation histories, dust content, gas distribution, and the growth of the supermassive black holes (SMBHs) residing in the nuclei of the merging galaxies.

Gravitational torques generated during the merger can drive gas from the outskirts of galaxies toward the inner regions \citep[e.g.][]{HopkinsSH06}. Gas might stall at the inner Lindblad resonance, typically 1 kpc from the nucleus, producing star-burst rings \citep[e.g.][]{ButaC96,PerezRamKP2000}. However, a chain of gravitational instabilities might be able to further transfer the gas toward the galactic nuclei, creating a reservoir that can fuel the SMBH \citep[e.g.][]{HopkinsQ10,EmsellemRB15}. The dark object will then reveal itself as an active galactic nucleus (AGN). Depending on the SMBH spin, mass, and the amount of gas available, the AGN might transfer mass and kinetic energy from the nucleus to the large-scale environment via outflows, regulating the star formation \citep[e.g.][]{CrenshawKG03MassLoss}. This interplay between host galaxy and AGN might originate the observed scaling relations between the SMBH and the bulge (e.g. \citealt{Merritt13}, Sec. 2.4.5, or \citealt{HeckmanB14}).

While the merger proceeds, activity switches on and off \citep[e.g.][]{WassenhoveVM12,HickoxMA14,SchawinskiKB15}. If both SMBHs are active, then the system, consisting of two interacting galaxies and two active SMBHs not yet gravitationally bound, becomes a ``dual AGN''. 

% ===========
% Figure SDSS with sources
% ===========
\begin{figure*}%[ht]
\center $
\begin{array}{ccc}
\includegraphics[trim=14cm 3cm 13cm 3cm, clip=true, scale=0.225]{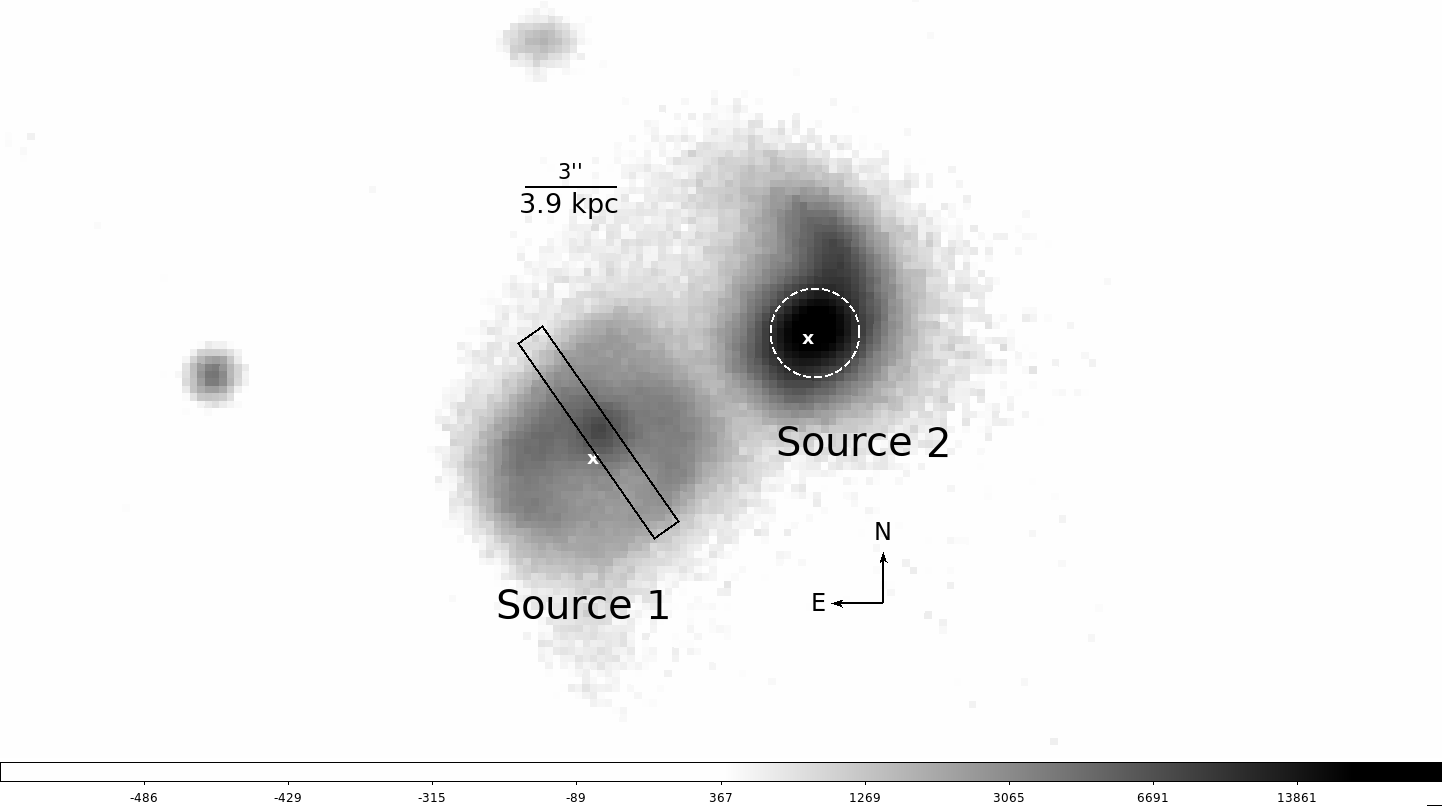}&
\includegraphics[trim=14cm 3cm 13cm 3cm, clip=true, scale=0.225]{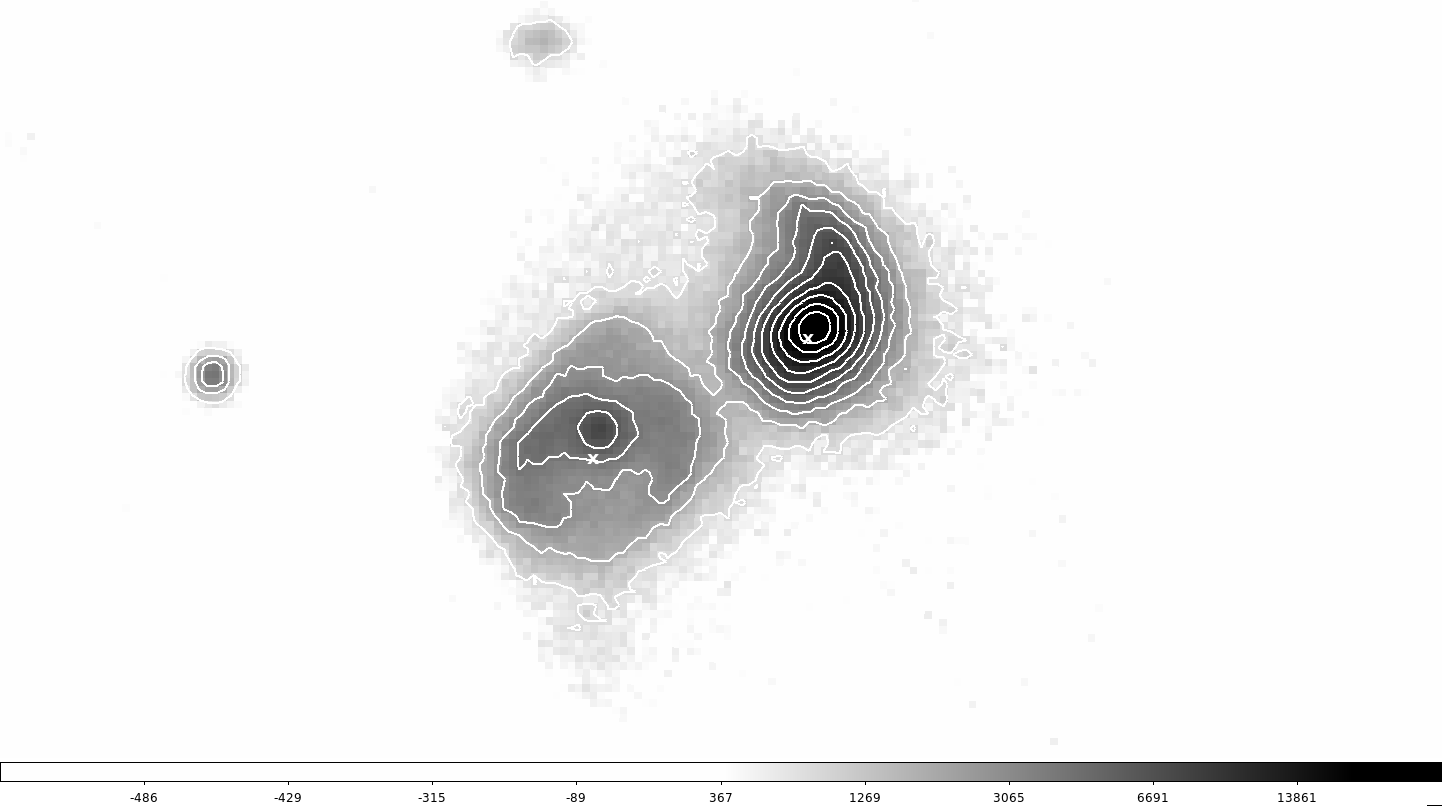} &
\includegraphics[trim=12cm 1.8cm 11cm 2cm, clip=true, scale=0.23]{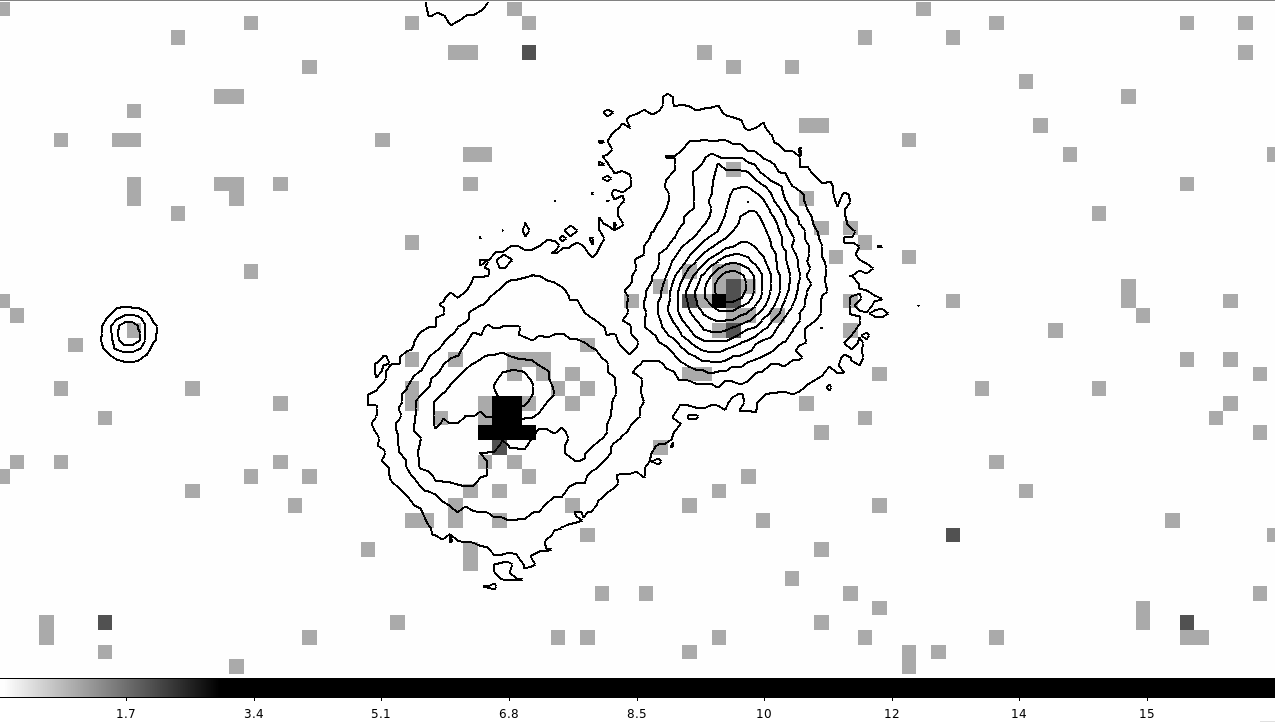}\\
\includegraphics[trim=12cm 1.8cm 11cm 2cm, clip=true, scale=0.23]{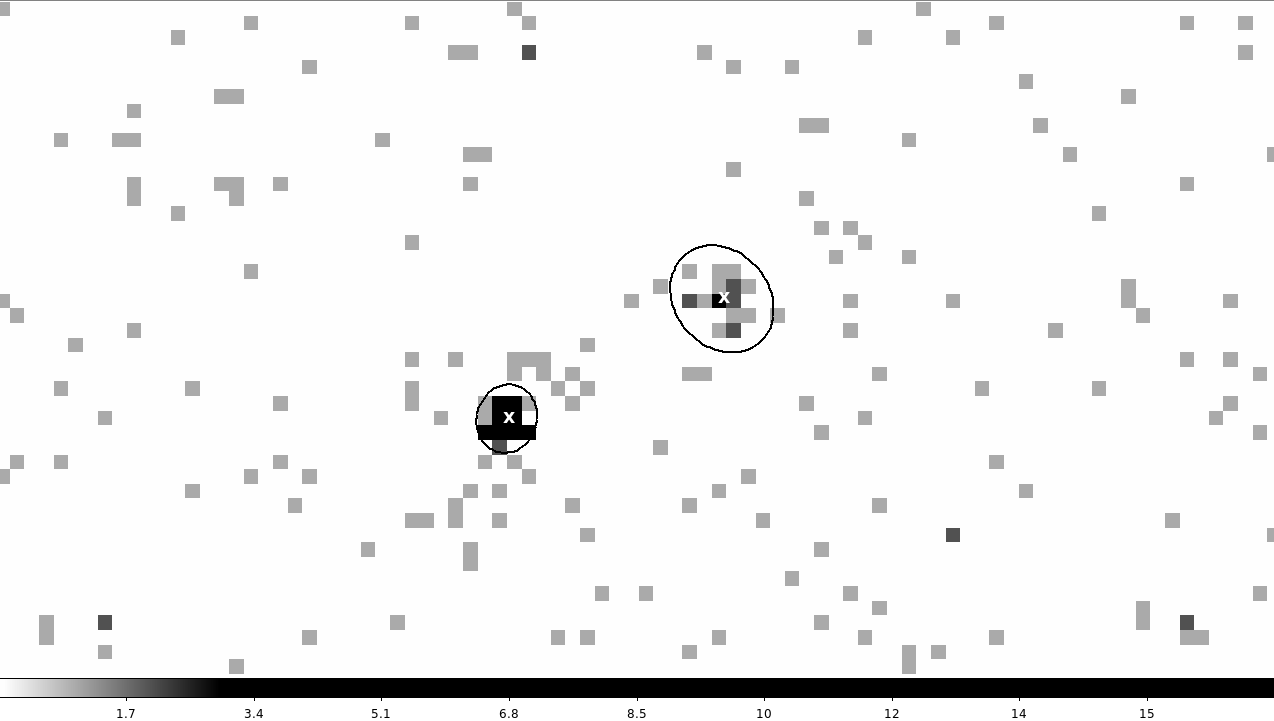}&
\includegraphics[trim=12cm 1.8cm 11cm 2cm, clip=true, scale=0.231]{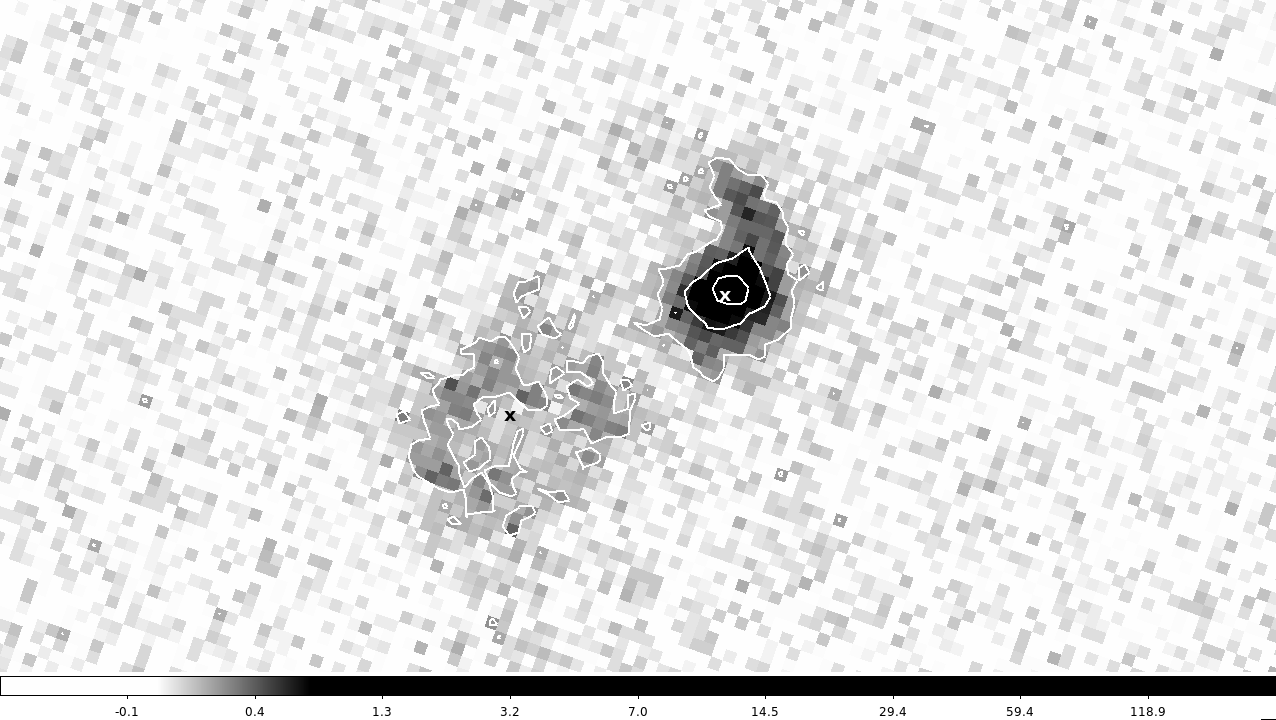} &
\includegraphics[trim=12cm 1.8cm 11cm 2cm, clip=true, scale=0.23]{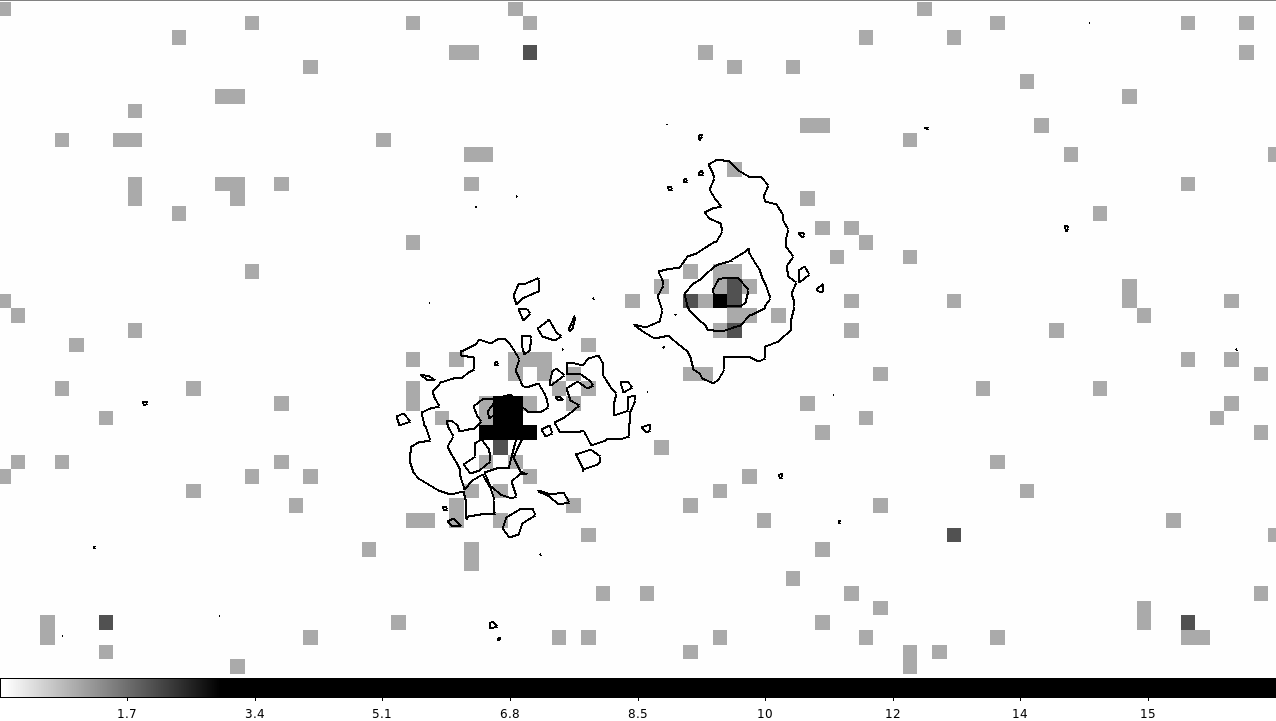}\\
\end{array}$
\caption{Top left: PanSTARRS-1 image of the two galaxies (\textit{r}-band, \citealt{ChambersMMF16}). The positions of the \co sources are indicated with a solid ``x'' of size $0\farcs4$ across. The dashed white circle represents position and size of the \textsc{sdss} fiber used to obtain the spectrum. The solid black box represents orientation and width of the \textsc{acam} slit. Top center: as in the left image, with a different colour scaling and \textit{asinh} contours overlaid to highlight the morphology of the galaxies. The white crosses represent the \co sources. Top right: X-ray \co image with optical contours overlaid. Bottom left: X-ray \co image with extraction regions (black ellipses) and their centroids (white ``x''). Bottom center: \textsc{sdss-dr13} image of the two galaxies (\textit{u}-band, \citealt{Albareti17}) with \textit{asinh} contours and \co sources indicated with ``x''. Bottom right:  X-ray \co image with \textit{u}-band \textsc{sdss} optical contours.}
\label{fig: xoptical}
\end{figure*}

In the mean time, the two SMBHs are believed to sink, because of dynamical friction, toward the minimum of the newly formed potential well. There the SMBHs will form a gravitationally bound system and, if dissipative mechanisms are efficient, their separation will decrease till the hardening of the orbit will be driven by the emission of gravitational waves \citep[e.g.][Ch.8]{Merritt13}. The two black holes will then merge, and asymmetries in the mass and spins of the SMBH-binary will result in the asymmetric emission of gravitational waves \citep[e.g.][]{Bek73}. This process will remove linear momentum from the new SMBH delivering it a kick, the amplitude depending on the properties and orbital configuration of the SMBH-binary \citep[e.g.][]{CampA,TM07,LoustoZ13}. The rarity of recoiling SMBHs (very few seemingly convincing cases have been found to date, e.g. \citealt{Civano2010,Civano12}), the small amplitude of AGNs spatially offset from their host nuclei \citep{lenaRMA14}, and the consistency between the SMBH occupation fraction at high and low redshift \citep{MerrittF01} suggests that recoils larger than a few 100 km s$^{-1}$ are rare, perhaps because the spins of the merging SMBHs are nearly aligned, due to prolonged accretion.

While the broad picture sketched above might be correct, solid observational support for all the steps is still far from being achieved. For example, the role of galactic mergers in the onset of AGN activity has long been debated with studies giving very different results even in the local universe (see \citealt{Alexander2012} for a review). Dual AGNs offer a snapshot of only one of the many phases taking place along the path to coalescence. Nevertheless, a thorough characterisation of a large sample of dual AGNs (with a range of physical separations, with a range of activity levels) will boost our understanding of a number of processes, such as the dynamics of stars, gas, and SMBH during the merger \citep[e.g.][]{EscalaBdV13,BlechaLN23,DottiMM15,KhanFM16}, the growth of SMBHs \citep[e.g.][]{CallegariKM11}, the merger rate of galaxies and, to a lesser measure, the merger rate of SMBHs (important quantities to test cosmological models and to make prediction for observable gravitational-waves events), and possibly we will gain insights on the expected electromagnetic counterparts of SMBH mergers \citep[e.g.][]{MilosavP05,Haiman17}. 

The majority of studies aiming to identify dual AGNs begin by selecting galaxies displaying double peaked emission lines, a spectral feature which may indicate the presence of two interacting narrow-line regions (NLR), and therefore two AGNs \citep[e.g.][]{Zhou04WZ,Wang09CH,ComerfordGS12,BarrowsSK13,ShiLC14}. However, it has long been known that double peaked emission lines could also originate from a broad range of phenomena, including unresolved rotation, non-gravitational motions in the ionised gas (i.e. inflows or outflows), structures in the NLR, interaction between jets and clouds, or one AGN illuminating two interacting galaxies \citep[e.g.][]{HeckmanMG84,KomossaZL08,Xu09K,CrenshawSK10,GabnyiFP17}. In recent years, integral field spectroscopy has shown at unprecedented level the complexity that the emission line profiles can display within active galaxies \citep[e.g.][]{SMullerEtAl11,LenaRSB14,DaviesGK17,NevinCMS18}. 
It is therefore clear that to confirm whether a candidate is a genuine dual AGN further insights are needed on the origin of the emission-line profiles \citep[e.g.][]{McGurkMM15,MullerSanchezCF16}, and on the nature of the galactic nuclei. Usually, this requires multi-wavelength follow up observations \citep[e.g.][]{KomossaBH03,FuZA11,KossMT11,MazzarellaIV12,LiuCS13,HainlineHC16,GabnyiAF16}.

Attempting to bypass the ambiguous origin of the optical emission line profiles, different approaches have been explored in the selection of dual AGNs, e.g.: \citet{KossMT12} started from the sample of ultra-hard X-ray selected AGNs in \citet{KossMV11} and looked for apparent companions within given radial and velocity bins. 
Recently, \citet{SatyapalSR17} targeted optically-obscured dual AGNs by selecting interacting galaxies displaying mid-infrared colours typical of AGNs.

In this paper we present X-ray and optical follow up for a candidate which we found serendipitously with still another method while searching for galaxies associated with single X-ray sources spatially offset from the optical hosts (possibly recoiling SMBHs). More specifically: in the Sloan Digital Sky Survey Data Release 7 (SDSS DR7, \citealt{AbazajianAMC09}) we searched for interacting galaxies characterised by a maximum separation of ten times the sum of their Petrosian radii \citep{Petrosian76}. Subsequently, we cross-correlated the sample with the XMM serendipitous source catalog (2XMMi, \citealt{WatsonSF09}) looking for X-ray sources within 1.5 arcmin from the SDSS positions.
Our approach yielded two candidates. The first one is the hyperluminous off-nuclear X-ray source presented in \citet[][a candidate recoiling SMBH]{JonkerTF10}. The second one is the candidate dual AGN presented in this work. The system consists on two galaxies (hereafter ``Source 1'', the Eastern, and ``Source 2'', the North-Western, Fig.\ref{fig: xoptical}) with a projected separation $\Delta r = 8\farcs1$; their SDSS IDs and coordinates are indicated in Table \ref{fig: xoptical}. 

This object was also listed among the AGN pairs selected by \citealt{LiuSS11} on the basis of the SDSS-DR7 emission line ratios.

Throughout the paper we assume the cosmological parameters H$_{0} = 70$ km s$^{-1}$ Mpc$^{-1}$, $\Omega_{m} = 0.3$, and $\Omega_{v} = 0.7$.

%%%%%%%%%%%%%%
% --------------------------------------------------------------------------------------------------------------------------------------------
\section{Observations and data reduction}
\label{sec_obs}
% --------------------------------------------------------------------------------------------------------------------------------------------
\subsection{X-ray}
% --------------------------------------------------------------------------------------------------------------------------------------------
X-ray imaging of the target was performed with the back-illuminated chip (S3) of the Advanced CCD Imaging Spectrometer (ACIS, \citealt{GarmireBF03}) of the \textit{Chandra X-ray Observatory} \citep{WeisskopfTVS00}. 
A 9.93 Ksec exposure was taken in ``timed exposure'' mode with telemetry formats set to ``very faint'' (observations ID: 12143).

We reduced the data using \textsc{ciao}\footnote{Chandra Interactive Analysis of Observations; \url{http://cxc.harvard.edu/ciao/}} \citep[v4.8,][]{FruscioneMDA06} and the calibration files \textsc{caldb v4.7.2}. The data reduction was performed with the application of the \textsc{ciao} tasks \textsc{chandra\_repro}, \textsc{wavdetect}, and \textsc{wcs\_update}.

The script \textsc{chandra\_repro} takes in input the raw data and generates a bad pixel file, an event file, and optimises spatial resolution via sub-pixel event repositioning \citep[e.g.][and references therein]{LiKP04}. Afterwards, we used \textsc{wavdetect} \citep{FreemanKR02} for source detection, and \textsc{wcs\_update} to align the \co and \textsc{SDSS} world coordinate system \textsc{(wcs)}. 

For \textsc{wavedetect} we adopted a significance threshold of $1\times10^{-7}$. In the energy range $0.3 - 0.7$ keV 10 sources were found in the image that was searched; that corresponds to the section of the S3 CCD delimited by the image coordinates (3650:4550, 3800:4275). Because of the faintness of most sources, we chose only one for the WCS update, hereafter \textsc{x1}, which implies that we ignored any uncertainty on the satellite roll-angle. The source x1 was detected by \textsc{wavedetect} with a significance of 18 (the second most significant source detected, after one of the target AGNs), and it corresponds to the star \sa, which also appears in \textsc{gaia} data release 1 \citep[\textsc{dr1,}][]{PerrymanBG01,BrownVP16}. Using the source position as given in \textsc{gaia-dr1}, we obtained a bore-sight \textsc{wcs} correction of $\Delta$(ra, dec) = ($-0\farcs29$, $0\farcs36$) $\pm\ 0\farcs08$. The uncertainty on the bore-sight correction is dominated by the uncertainty in localising the X-ray source on the \textsc{acis-s3 ccd} using the \co data; the uncertainty on the position of the reference star, as measured with \textsc{gaia}, is of order $1\ mas$. We did not account for the star proper motion.

Proper motion and position are available for the star in UCAC4 \citep{ZachariasFG13}, however we deemed the derived correction to be less accurate: when propagating the UCAC4 position of the star with its proper motion to the epoch of Gaia DR1, we obtained an offset of $0\farcs3$ and $0\farcs1$ in \textsc{ra} and \textsc{dec} between this inferred position and the one in Gaia.

% =========
% Table summary
% =========
\begin{table}
\begin{center}
\caption{X-ray sources location. (1) Source ID; (2) SDSS-DR7 ID; (3 - 6) coordinates and centroid uncertainties for the \co sources after bore-sight correction. Uncertainties are in arcseconds and do not take into account the $0\farcs08$ systematic uncertainty introduced by the bore-sight correction.}  
\scalebox{0.8}{
\begin{tabular}{l lllllll}
\hline
	 Source	&  	SDSS name			& R.A.			& R.A. err		& Dec. 			& Dec. err\\
	 (1)		&	(2)					& (3)				& (4)			& (5) 			& (6)\\
\hline
& &    \\
1			& J085312.85$+$162616.0	& $133\fdg30362$	& 0\farcs04	& $16\fdg43756$ 	&	0\farcs06\\
2			& J085312.34$+$162619.4	& $133\fdg30153$	& 0\farcs2		& $16\fdg43868$ 	&	0\farcs1\\
	
& &    \\
\hline
\end{tabular}}
\label{tab: coo}
\end{center}
\end{table}

% ===========
% Observed x-ray spectra
% ===========
\begin{figure*}%[th]
\center $
\begin{array}{cc}
\includegraphics[trim=1cm 1cm 2cm 2cm, clip=true, scale=0.365]{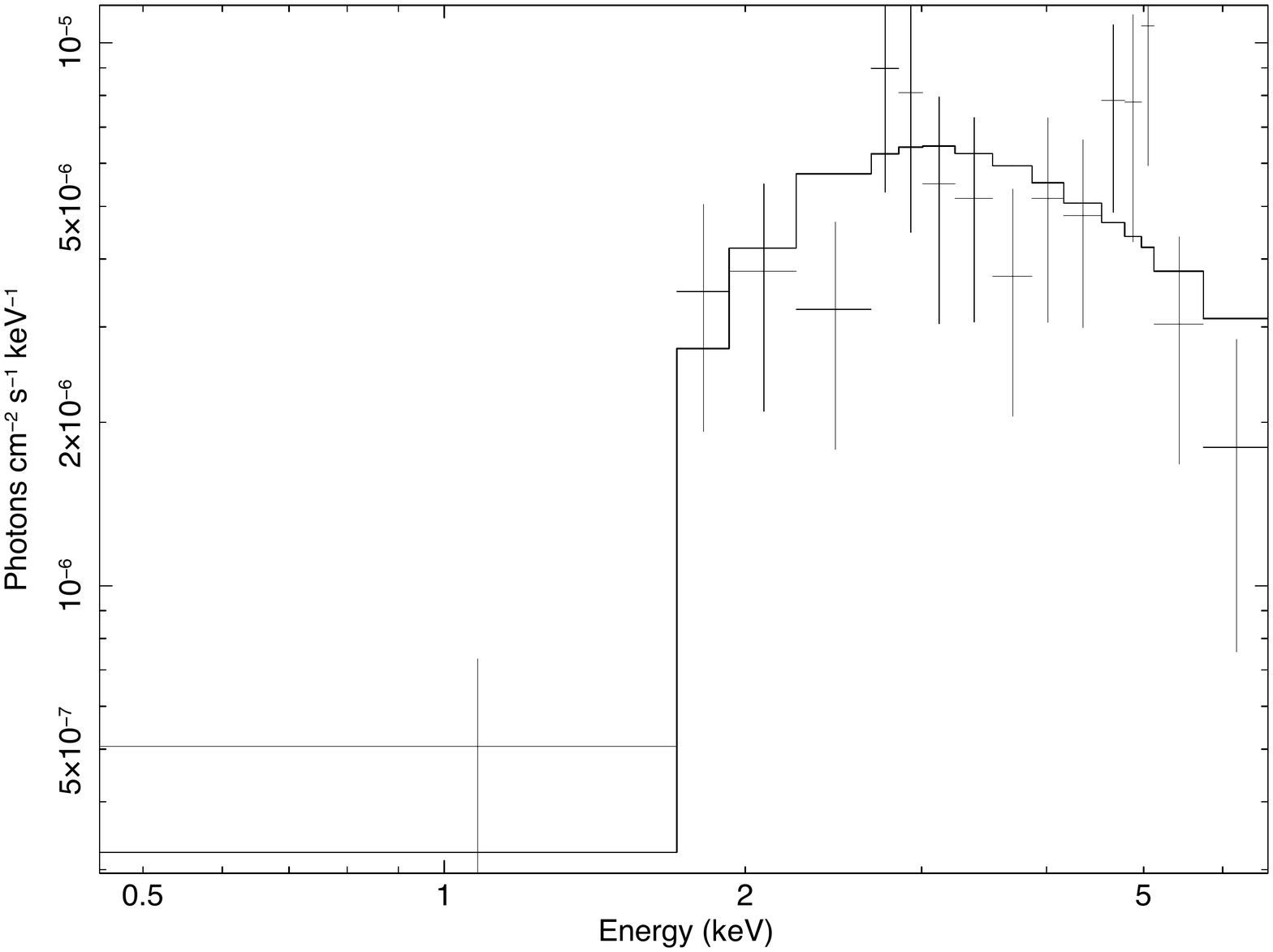} & \includegraphics[trim=1cm 1cm 2cm 2cm, clip=true, scale=0.365]{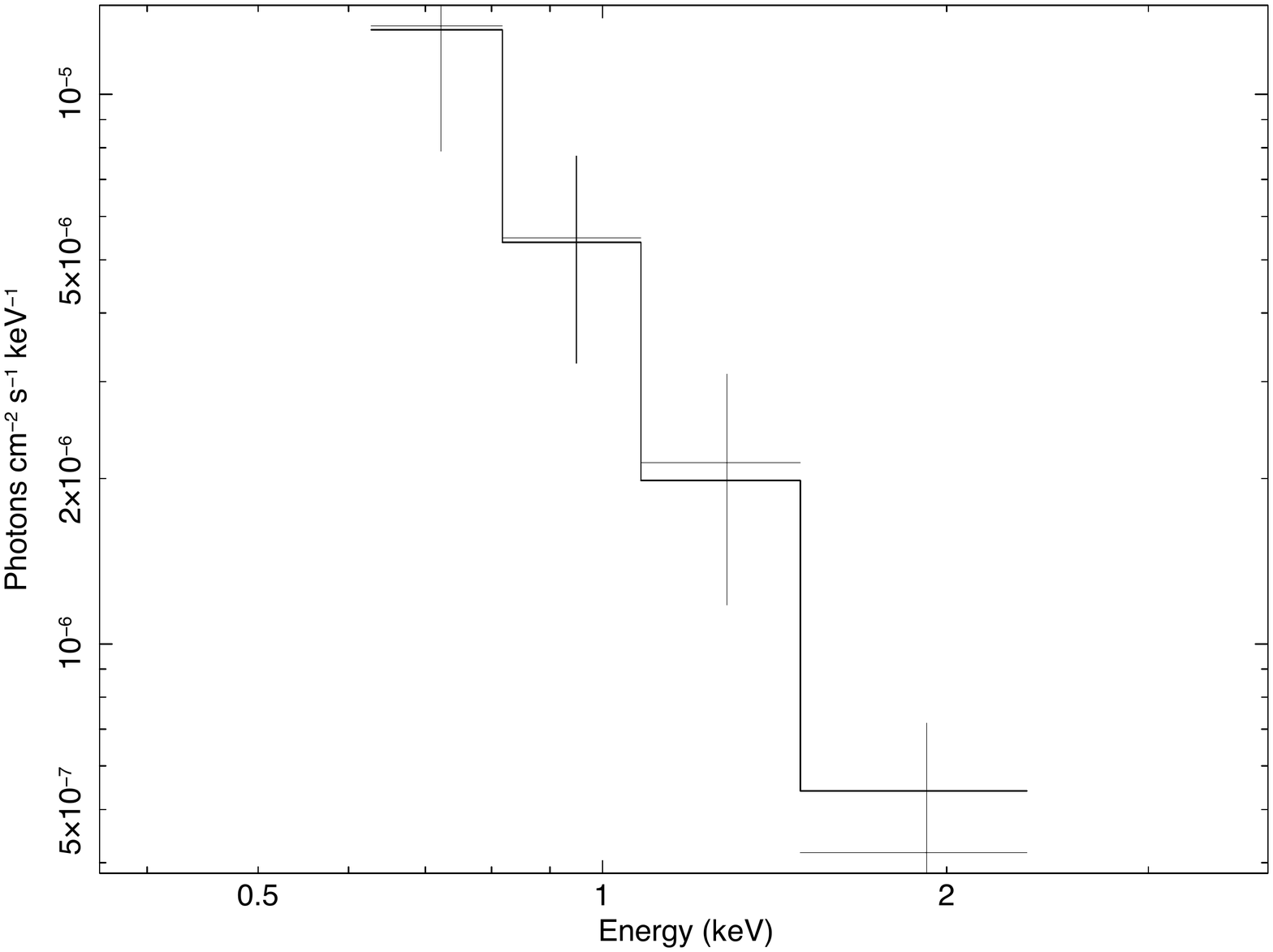}
\end{array}$
\caption{Observed \co spectra, and \textsc{xspec} fit, for Source 1 and 2.}
\label{fig: xspec}
\end{figure*}

% =========
% Table summary
% =========
\begin{table*}%[t]
\begin{center}
\caption{Best fitting parameters for the X-ray spectra. (1) Source ID; (2) column density, units $cm^{-2}$; (3) power-law index; (4) observed flux; (5, 6) unabsorbed flux, untis: \textit{erg s$^{-1}$ cm$^{-2}$}; (7) observed luminosity; (8, 9) unabsorbed luminosity, untis: \textit{erg s$^{-1}$}. Uncertainties represent $68\%$ confidence levels. The main source of uncertainty is the column density. The distances assumed to derive the fluxes, $D_{L,1} \approx 291$ Mpc, and $D_{L,2} \approx 288$ Mpc, have been derived in Sec. \ref{subsec: interact}. Observed fluxes in the band 2 - 10 keV are consistent with the unabsorbed fluxes. $^{\dagger}$The flux in the 2-10 keV range for Source 2 was extrapolated from the fit; no emission was observed above 2 keV.}  
\begin{tabular}{l lllllllll}
\hline
	 Source	& $n_{H}$				&  	 $\Gamma$		&	F$^{\mathrm{\ obs}}_{0.3 - 7\ \mathrm{keV}}$	& F$_{0.3 - 7\ \mathrm{keV}}$			&	F$_{2 - 10 \ \mathrm{keV}}$		& L$^{\mathrm{\ obs}}_{0.3 - 7\ \mathrm{keV}}$		& L$_{0.3 - 7\ \mathrm{keV}}$	& L$_{2 - 10\ \mathrm{keV}}$			& \\
	 (1)		& (2)					&	(3)				&	(4)									& 	(5)							&	(6)							& (7) 									& 	(8)					& (9)\\
\hline
& &    \\
1			& $5 \times 10^{22}$		& 1.9					&	$1.6^{+0.2}_{-0.12} \times 10^{-13}$				& $5.7_{-0.12}^{+0.2} \times 10^{-13}$ 	& $2.2_{-0.3}^{+0.2} \times 10^{-13}$	& $1.6^{+0.2}_{-0.1} \times 10^{42}$				&$5.8_{-0.1}^{+0.2} \times 10^{42}$	& $2.2_{-0.3}^{+0.2} \times 10^{42}$	& \\
& &    \\
2			& $2 \times 10^{20}$		& $3.4^{+0.8}_{-0.7}$	&	$2.3^{+2}_{-1} \times 10^{-14}$				& $2.8_{-1}^{+2} \times 10^{-14}$ 		& $1.7_{-1.2}^{+1.3} \times 10^{-15}$	& $2.3^{+2}_{-1} \times 10^{41}$				&$2.8_{-1}^{+2} \times 10^{41}$	& $1.7_{-1.2}^{+1.3} \times 10^{40}$	& $\dagger$\\
	
& &    \\
\hline
\end{tabular}
\label{tab: xray}
\end{center}
\end{table*}

% --------------------------------------------------------------------------------------------------------------------------------------------
\subsection{Optical}
SDSS spectra are available for both galaxies. However, the fiber used to observe Source 1 was not centered on the nucleus of the galaxy, which is the region which we want to probe. For this reason we obtained a new spectrum using ACAM \citep{BennDA08} on the William Herschel Telescope.

Observations were carried out on April 12 2017. The slit was oriented at the parallactic angle, the width was 1$^{\prime\prime}$, and the seeing varied between 0\farcs9 and 1\farcs2. We obtained two exposures of 900 seconds each, centred on the nucleus, using the grism V400. The nominal instrumental dispersion is $\Delta \lambda \approx 3.3$ \AA\textit{/pixel}. For a slit width of 1$^{\prime\prime}$ the predicted resolving power is R$=$430 at 5650 $\mathrm{\AA}$, or $\Delta \lambda = 13.1$\AA\ and $\sigma = 296$ km s$^{-1}$, which is consistent with the value that we measured from the [OI]$\lambda$5577 sky-emission line (this value is likely an over-estimate as we measured smaller velocity dispersions from the fit of the spectrum, Sec. \ref{subs: fitoptical}). To perform flux calibration, we observed the standard star BD+08d2015 (spectral type G2V) immediately before the target.

Data reduction was performed using \textsc{iraf}\footnote{\textsc{iraf} (Image Reduction and Analysis Facility), is a general purpose software system for the reduction and analysis of astronomical data. It is distributed by the National Optical Astronomy Observatory, which is operated by the Association of Universities for Research in Astronomy (AURA) under a cooperative agreement with the National Science Foundation.}. The data were corrected for the bias, flatfielded, calibrated in wavelength, and background subtracted. Variance-weighted spectra were extracted using an aperture of diameter 3\farcs5 centred on the nucleus, a heliocentric correction of -28.29 km s$^{-1}$ was applied, fluxes were calibrated, and the two exposures combined into an average spectrum.

From the \textsc{rms} residual of the \textsc{iraf} task \textsc{reidentify}, we estimated that the wavelength calibration is accurate to within 0.5 $\mathrm{\AA}$. Before flux calibration, a shift of 14 $\mathrm{\AA}$ was applied to the wavelength axis to account for a systematic shift in the wavelength calibration zero-point; the bias is likely due to the fact that the arc was not obtained immediately before/after the science exposure; a shift of this amplitude is typical for \textsc{acam}\footnote{\url{http://www.ing.iac.es/Astronomy/instruments/acam/flexuretests.html}}).The amplitude of the shift was estimated from the wavelength of the sky lines [OI]$\lambda$5577.3, NaI$\lambda$5890, and [OI]$\lambda\lambda$6300.3,6363.8. The correction is estimated to be accurate to within 0.1 $\mathrm{\AA}$.

% ================================================================================
\section{Data analysis and results}  \label{sec_result}
% --------------------------------------------------------------------------------------------------------------------------------------------
\subsection{X-ray sources and galactic nuclei} \label{subsec: xrayf}
% --------------------------------------------------------------------------------------------------------------------------------------------
As specified in Section \ref{sec_obs}, we aligned the coordinates of the \co data with the International Coordinate Reference System (\textsc{icrs}) using a \textsc{gaia-dr1} source that was detected in X-rays on the same CCD as the target under consideration. Applying the coordinate correction allowed a more accurate comparison between optical and X-ray data showing that two X-ray sources are approximately coincident with the optical nuclei of the two galaxies, Fig.\ref{fig: xoptical}. The spatial offset between the X-ray source and the optical centroid is larger for Source 1; while the position of the X-ray centroid is consistent with the centroids derived from bluer optical images ($u$ and $g$ bands), the spatial offset increases at longer wavelengths, up to about 0\farcs8.
 
The corrected coordinates for the X-ray sources are given in Table \ref{tab: coo}. Using \textsc{wavdetect} we counted $91 \pm 10$ X-ray photons for Source 1, with a significance of 36, and $21 \pm 5$ photons for Source 2, with a significance of 8, and with uncertainties estimated assuming Poisson statistics. Counts were extracted using \textsc{wavdetect} from the elliptical regions shown in Fig.\ref{fig: xoptical}; they were centred on the X-ray sources with axes and position angles equal to 1\farcs66, 1\farcs77, -15$^{\circ}$ East of North, for Source 1, and 2\farcs45, 1\farcs55, 40$^{\circ}$ East of North, for Source 2. Using the \textsc{ciao} script \textit{fluximage}, we obtained consistent source counts in the broad energy band $0.5 - 7$ keV.

We extracted the background from a circular region of radius 40$^{\prime\prime}$ in an area of the \textsc{s3 ccd} devoid of sources. The background amounts to $< 1$ photon, which was subtracted from the counts given above.

To determine whether the X-ray emission is consistent with an AGN origin, we modelled the spectra of the sources, which were extracted from the regions specified above.
To model the spectra we used \textsc{xspec}\footnote{\url{https://heasarc.gsfc.nasa.gov/xanadu/xspec/}} \citep[v.12.9,][]{Arnaud96}, and we adopted Cash statistics \citep{Cash79} to fit an absorbed power-law. Toward this end we used the \textsc{xspec} function \textit{pegpwr $\times$ phabs}, that is:

\begin{align}
f(E) = KE^{-\alpha}exp[-n_{H}\sigma(E)],
\end{align}

\noindent where \textit{E} is the energy, $K$ the unabsorbed flux, $\sigma$(E) is the photo-electric cross-section (not including Thomson scattering), $n_{H}$ is the equivalent hydrogen column (in units of $10^{22}$ atoms cm$^{-2}$), $\alpha = \Gamma - 1$ is the power-law index, with $\Gamma$ the ``photon index''.

From the fit we derived the power-law index, and the unabsorbed flux. For Source 2 the hydrogen column density was fixed to $n_{H} = 2 \times 10^{20}$ cm$^{-2}$ \citep[i.e. the Galactic value,][]{KalberlaBH05}. As a similar assumption would produce a negative power law for Source 1, in this second case we determined the value of $n_{H}$ which would produce a power-law index $\Gamma = 1.9$, which is typical for AGNs \citep[e.g.][]{RisalitiElvis04}. This resulted in $n_{H} = 4.7 \times 10^{22}$ cm$^{-2}$. 

Results of the fit are summarised in Table \ref{tab: xray}. Spectra for the two sources along with the fits are presented in Fig.\ref{fig: xspec}. The spectra have been rebinned for display purposes only.

% ======================================
\begin{figure*}%[th]
\begin{center}$
\begin{array}{cc}
\includegraphics[trim=0cm 0cm 0cm 0cm, clip=true, width=0.4\textwidth]{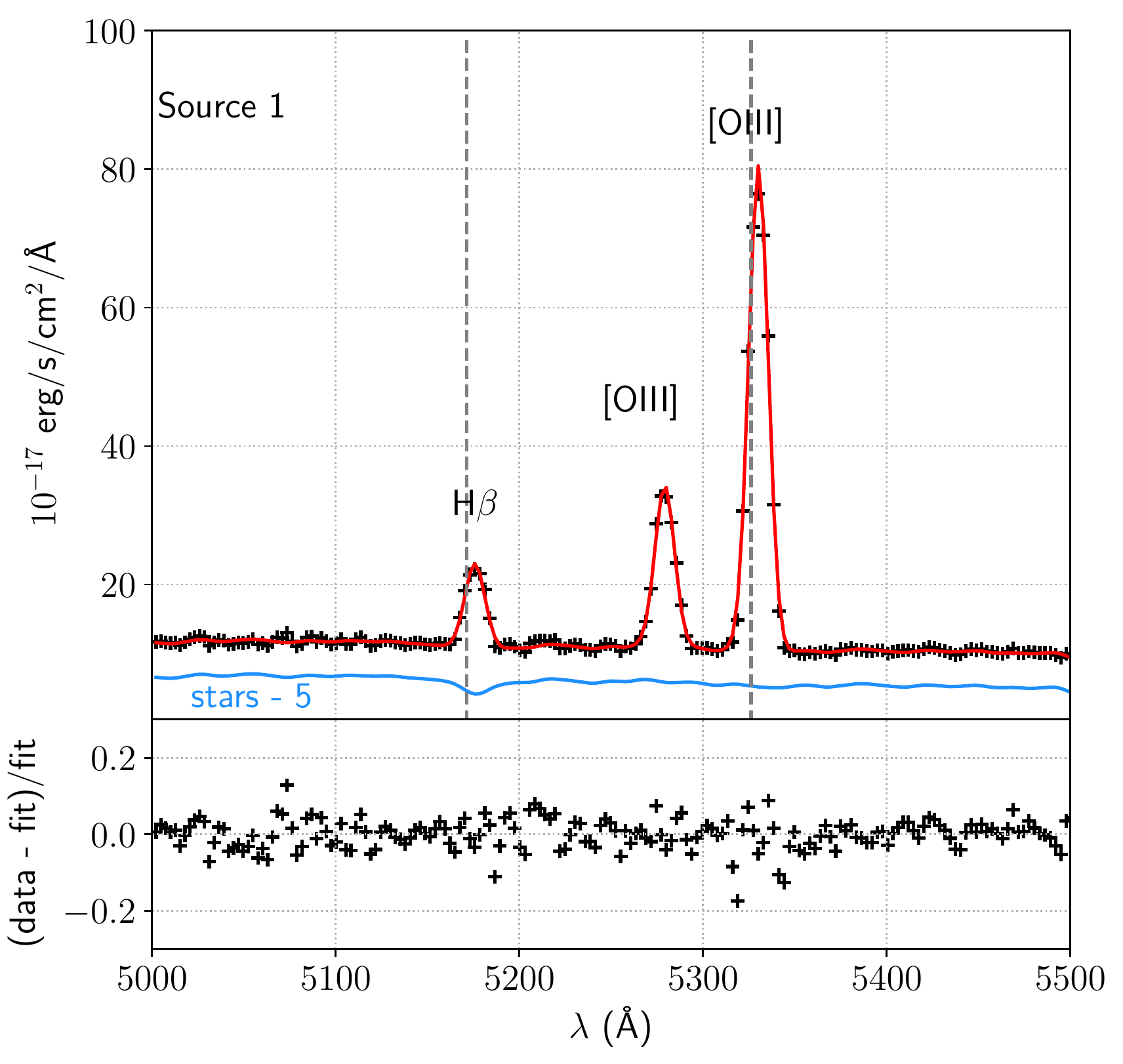}	& \includegraphics[trim=0cm 0cm 0cm 0cm, clip=true, width=0.4\textwidth]{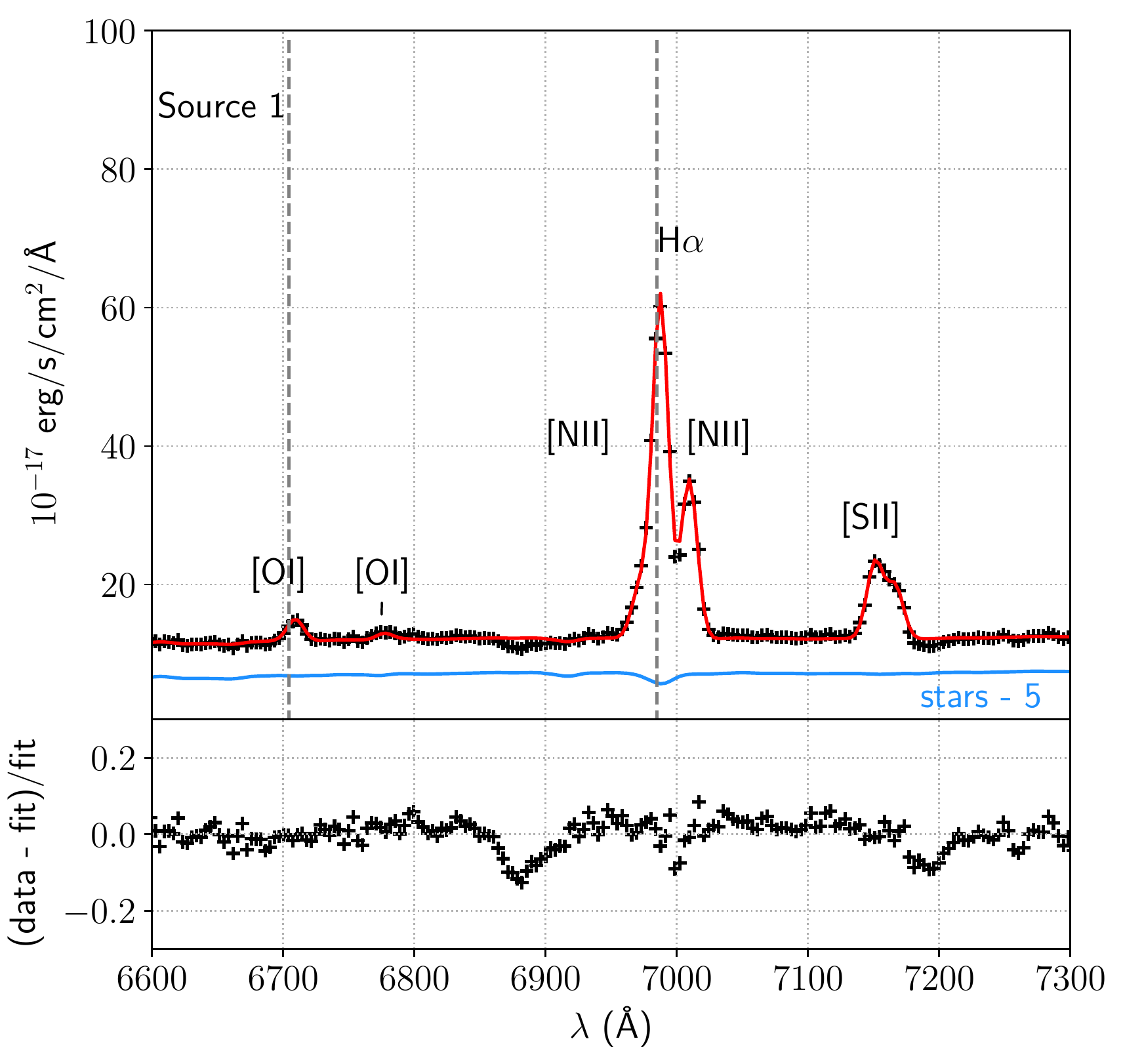} \\
\includegraphics[trim=0cm 0cm 0cm 0cm, clip=true, width=0.4\textwidth]{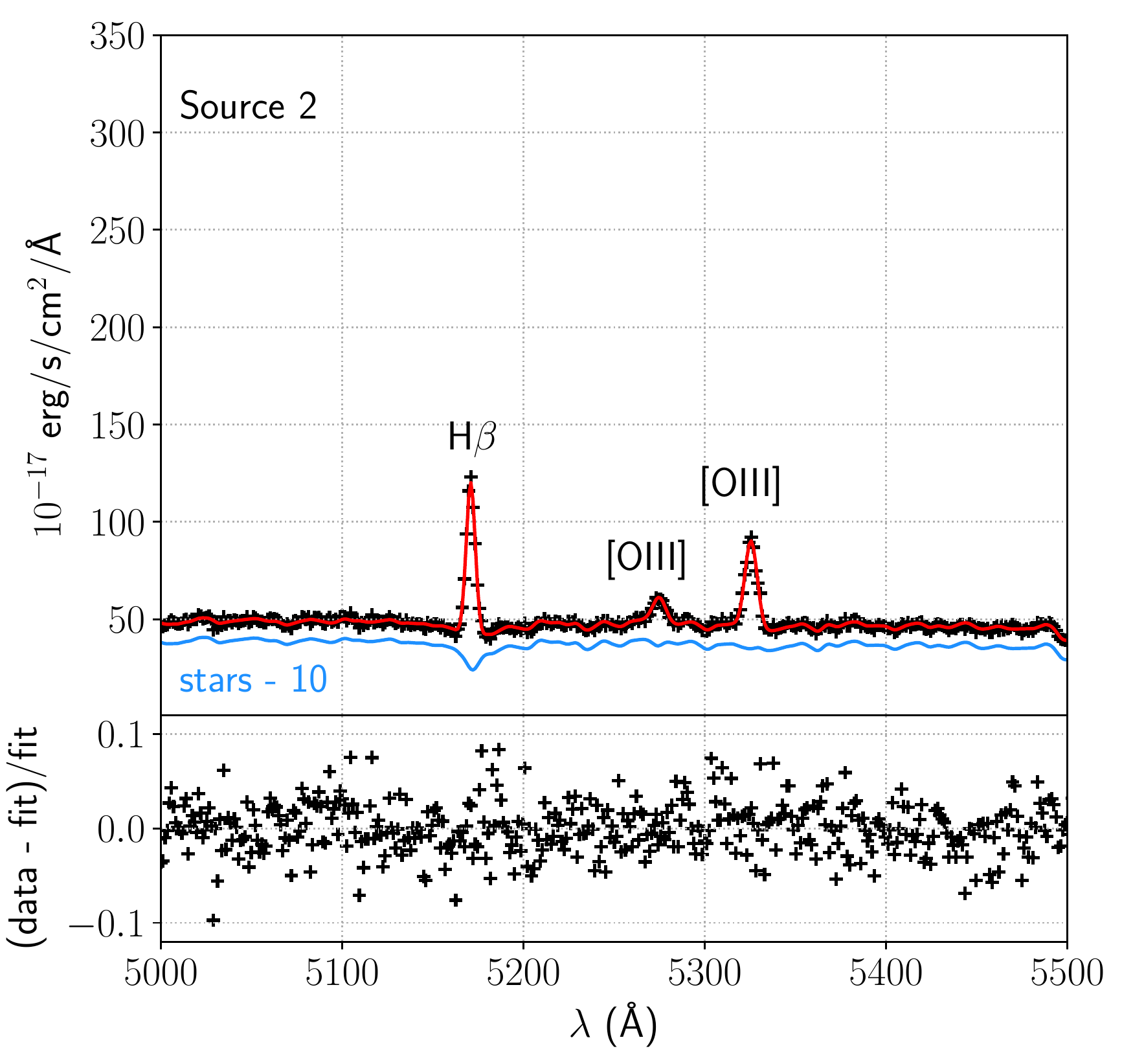}	& \includegraphics[trim=0cm 0cm 0cm 0cm, clip=true, width=0.4\textwidth]{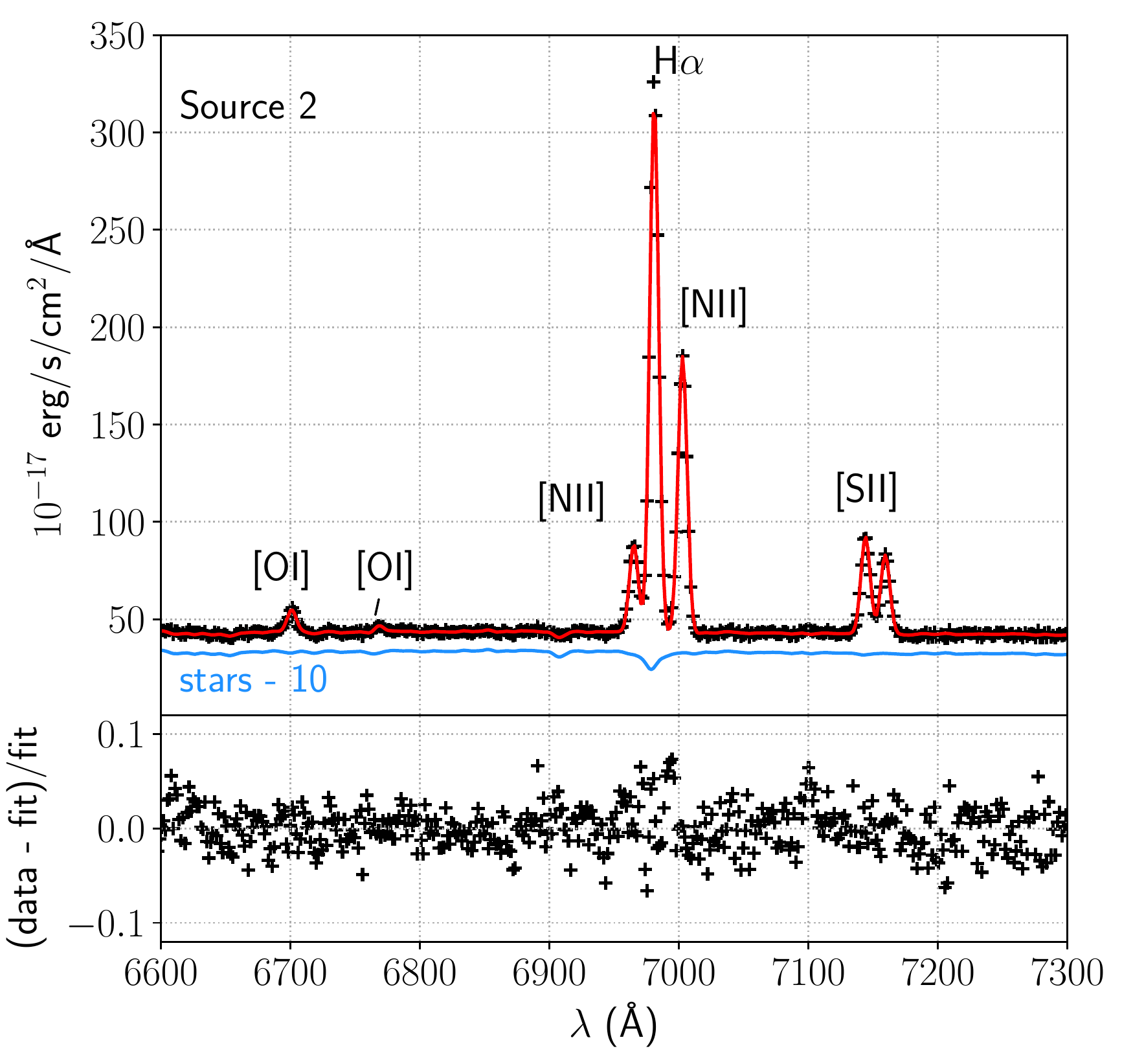} \\
\end{array}$
\end{center}
\caption{Top: ACAM spectrum for Source 1. Bottom: SDSS-DR12 spectrum for Source 2 (spectrum ID: 2431-53818-0452). Data-points are plotted as black `+', the fit (including gas emission lines and the stellar continuum) is shown in red, the fitted stellar continuum is shown in blue and offset downward for clarity. To facilitate visual comparison, vertical dashed grey lines in the top panels indicate the peak of the corresponding emission lines derived for Source 2.}
\label{fig: spec_s2}
\end{figure*} 

% --------------------------------------------------------------------------------------------------------------------------------------------
\subsection{Fit of the optical spectrum}
\label{subs: fitoptical}
% --------------------------------------------------------------------------------------------------------------------------------------------
The optical spectra of the two galaxies show a number of emission and absorption lines. We used pPxf v6.6.1 \citep{CappellariEmsellem04,Cappellari17} to fit the spectral region 5000 - 7500 \AA, which includes the emission lines H$\beta$, [OIII]$\lambda$4959,5007, [OI]$\lambda$6300, H$\alpha$, [NII]$\lambda$ and [SII]6716,6731, and several stellar absorption features.  
The fit of the spectra was performed in two steps: initially, we fitted the absorption features (i.e. the stellar kinematics); afterwards, we held fix the parameters of the stellar kinematics, as recovered from the previous step, and we fitted the full spectrum, including the emission lines. 

More specifically, to fit the stellar kinematics we masked out any emission line, and the Na I $\lambda\lambda5890,5896$ absorption lines, which are affected by the interstellar medium of the host galaxy. To reproduce the absorption line profiles, we used a subset of the \textsc{miles} stellar templates \citep{SanchezB06}, and we modelled the line profiles of the \textsc{sdss} spectrum as Gauss-Hermite (GH) polynomials; as the \textsc{sdss} wavelength calibration was performed using ``in vacuum'' values, while the \textsc{miles} templates used ``in air'' values, we used the \textit{vac\_to\_air()} function provided with the pPxf package to corrected the wavelength axis of the \textsc{sdss} spectrum to ``in air'' values. Due to the lower resolution of the \textsc{acam} data, in this second case we assumed a Gaussian profile.

To fit the emission lines we used four spectral components, each one modelled as a Gaussian. In particular, we used a spectral component to fit the H$\beta$ emission line, one for the [OIII] doublet, one for the [OI] doublet, and a fourth one to fit H$\alpha$, [NII], and [SII]. 

Default values were adopted for the pPxf parameters \textit{bias} and \textit{regul\_err}, that is $0.386$ and $0.0036$ respectively, for the \textsc{sdss} spectrum; the same value was adopted for \textit{regul\_err} in the fit of the \textsc{acam} spectrum, while the \textit{bias} value was irrelevant because a Gaussian profile was adopted. The \textit{noise} array given to pPxf was the inverse square root of the inverse variance array. The fit of the \textsc{sdss} spectrum took into account the intrinsic dispersion of every pixel using the information provided within the \textsc{sdss} data set in the array \textit{wdisp}. The resulting velocity dispersions were therefore corrected for instrumental broadening by pPxf itself. As a dispersion array for \textsc{acam} was not available, we performed the fit without any assumption on the instrumental dispersion, obtaining ``observed'' velocity dispersions, i.e. not corrected for instrumental broadening. 
Results of the fit (redshift, velocity dispersion, fluxes, and GH coefficients) are shown in Table \ref{tab: eml}, and in Fig.\ref{fig: spec_s2}. 

We assessed the robustness of our solution against different values of the initial guesses (for redshift and velocity dispersion), and we checked how different values of the parameters \textit{bias} and \textit{regul\_err} would affect the result. The solution showed negligible variations through the experiment. As expected, the most important variations were observed for the GH coefficients, which are biased toward larger or smaller values by the parameter \textit{bias}. This variation in the GH coefficients produced a variation of 1 km s$^{-1}$ in the stellar velocity dispersion.

To estimate the statistical uncertainties due to the fitting procedure, we performed a Monte Carlo simulation: to the observed spectrum we added Gaussian noise with mean zero and standard deviation equal to the value of the noise array at that specific wavelength. We fitted the resulting spectrum, and we repeated the process 100 times. From each fit we derived redshift, velocity dispersion, and flux (for the emission lines). The standard deviation of the parameters distribution was assumed to represent the wanted uncertainty, and it was derived using the \textsc{numpy} function \textit{std()} with the parameters \textit{ddof()} and \textit{dtype} set to zero and \textit{np.float64}, respectively. Results are indicated in Table \ref{tab: eml}.  

% ===========
% Color map g - y
% ===========
\begin{figure*}%[th]
\center $
\begin{array}{cc}
\includegraphics[trim=0cm 0cm 0cm 0cm, clip=true, scale=0.7]{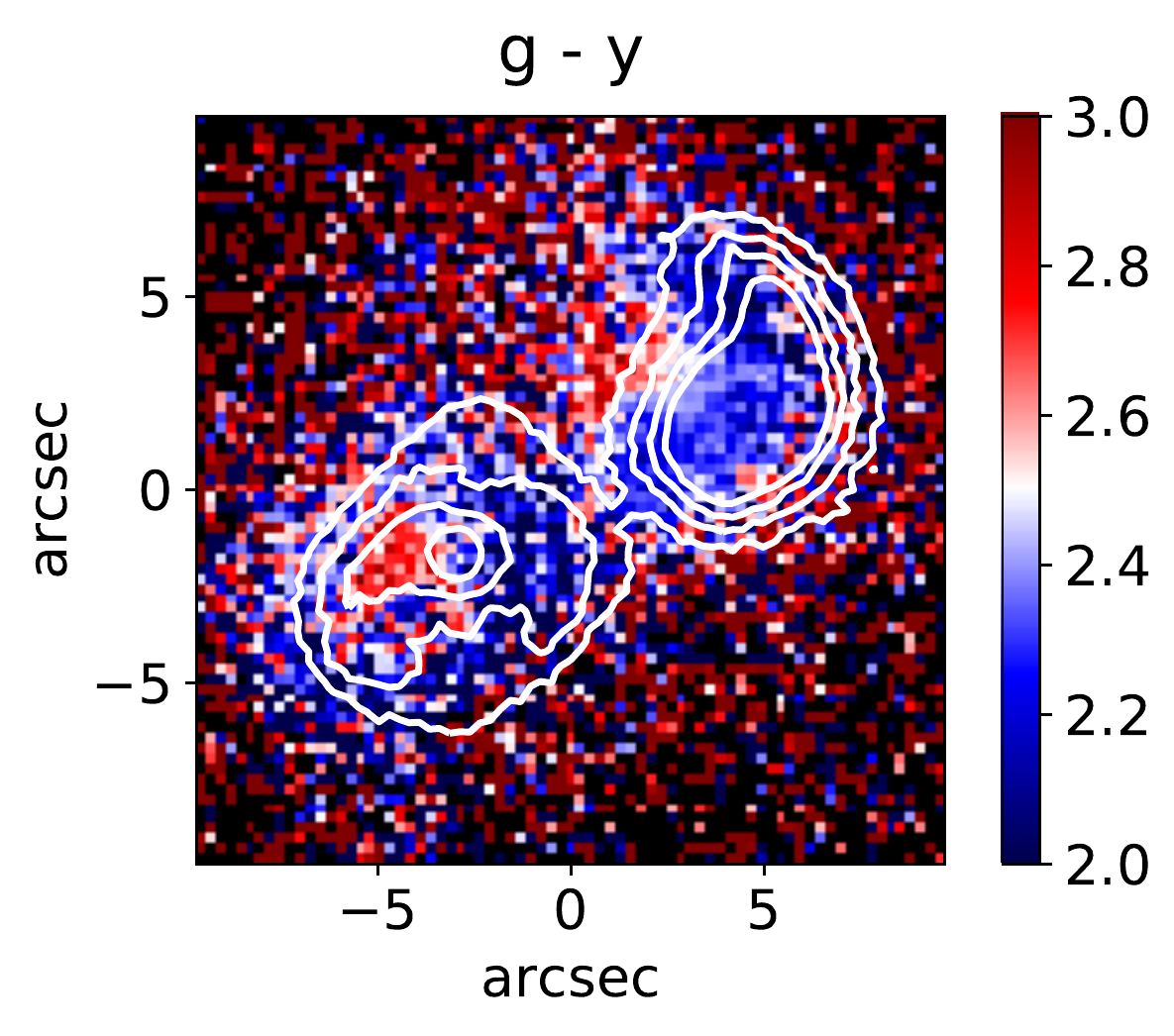} & \includegraphics[trim=-0.5cm -0.8cm 0cm 0cm, clip=true, scale=0.47]{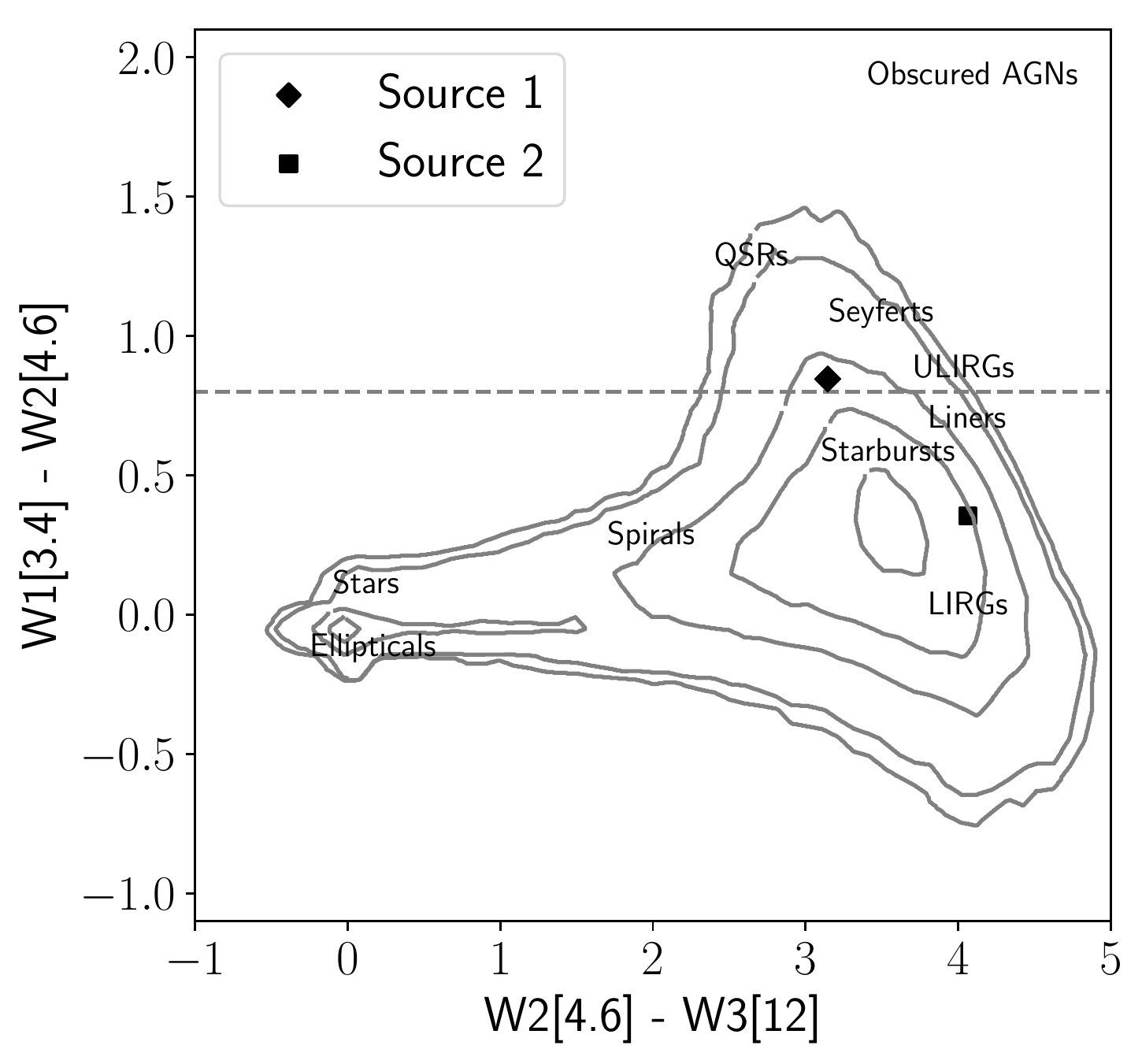}
\end{array}$
\caption{Left: optical colour map (\textit{g} - \textit{y}) with \textit{r}-band contours to highlight the most prominent photometric features. The innermost contours have been removed from Source 2 for clarity. Right: \textsc{wise} colours overlaid on the colour-colour plot of \citet{MassaroAA11}. The contours represent isodensity curves for approximately 450000 \textsc{wise} sources. The location of different classes of objects is annotated. The horizontal line at \textsc{w1-w2} = 0.8 indicates the threshold proposed by \citet{SternAB12} between \textsc{agn} candidates and non-active galaxies. Uncertainties are smaller than the symbol size.}
\label{fig: gmy}
\end{figure*}

% --------------------------------------------------------------------------------------------------------------------------------------------
\subsection{Electron density}
% --------------------------------------------------------------------------------------------------------------------------------------------
We estimated the electron density (n$_{e}$) using the [SII]$\lambda$6716/6731 emission line ratios \citep[e.g.][Sec. 5.6]{OsterbrockF06_book}.

From the flux values in Table \ref{tab: eml}, we derived the ratios $r_{1} = 1.5 \pm 0.1$ and $r_{2} = 1.24 \pm 0.05$. We assumed a temperature $T = 10^{4}$ K, and we used the \textsc{iraf} task \textit{temden} \citep{DeRobertisDH87,ShawD95} to derive the corresponding densities.

The line ratio in Source 1 falls above the maximum theoretical limit, which corresponds to the low-density limit; we infer n$_{1,e} < 27$ e$^{-}$ cm$^{-3}$. 
In Source 2 we find n$_{2,e} = 190 \pm 60$ e$^{-}$ cm$^{-3}$. 

Star forming galaxies tend to have lower densities than Seyferts and \textsc{liner}s \citep{Ho96}. For a sample of local ($0.04 < z < 0.1$) star-forming galaxies, \citealt{SandersSK16} obtained a median [SII] line ratio of 1.41 (corresponding to $n_{e} = $ 26 e$^{-}$ cm$^{-3}$ with their prescription), and they observed line ratios in the range 0.9 $<$ r $<$ 2.1 (corresponding to densities n$_{e} \lesssim 600$ e$^{-}$ cm$^{-3}$). In the nuclei of local Seyferts, electron densities span the whole sensitivity range of the [SII] doublet (n$_{e} < 1500$ e$^{-}$ cm$^{-3}$), with typical values of a few hundreds electrons per cm$^{-3}$ \citep[e.g.][]{BennertJK06,LenaRSB16,BrumRSB17}. For the Palomar Survey of Nearby Galaxies sample, \citealt{Ho96} quoted a median value for Seyfert galaxies of n$_{e}$ = 290 e$^{-}$ cm$^{-3}$.

% --------------------------------------------------------------------------------------------------------------------------------------------
\subsection{Optical and mid-infrared colours}
\label{subsec: col}
% --------------------------------------------------------------------------------------------------------------------------------------------
An optical colour map of the system was produced using the \textit{g}- and \textit{y}-band PanSTARRS images \citep{ChambersMMF16}. 
The colour map was computed as:

\begin{align}\nonumber
g - y = -2.5 log_{10}\frac{f_{g}}{f_{y}} + m_{0,g} - m_{0,y},
\end{align}

\noindent where $f = counts/exposure\ time$ is the linear pixel flux (as obtained for the cutout images from the PanSTARRS archive\footnote{\url{http://ps1images.stsci.edu/cgi-bin/ps1cutouts}}); $m_{0,g} $ and $m_{0,y}$ are the magnitude zero points, which in the cutout images are both set to 25. 
 The resulting map is shown in Fig.\ref{fig: gmy}.

There is tentative evidence for resolved structures: the bar-like feature in Source 1 shows redder colours eastward of the nucleus. Source 2 also shows a gradient with bluer colours, on average, in the tail.

To derive the mid-infrared colours we used photometric data from the ``All \textsc{wise} Source Catalog''\footnote{\url{http://irsa.ipac.caltech.edu/cgi-bin/Gator/nph-scan?submit=Select&projshort=WISE}}. Results are indicated in the right panel fo Fig.\ref{fig: gmy}, where the colours for Source 1 and Source 2 are plotted on the colour-colour diagram of \citet{MassaroAA11}. Source 1 falls near the locus of Seyfert galaxies, and above the threshold proposed by \citet{SternAB12} for the identification of candidate \textsc{agn}s (i.e. \textsc{w1-w2} $\geq$ 0.8). Source 2 falls in between the locus of Starbursts and \textsc{lirg}s.

% --------------------------------------------------------------------------------------------------------------------------------------------
\subsection{2D decomposition}\label{subsec: galfit}
% --------------------------------------------------------------------------------------------------------------------------------------------
We used \textsc{galfit} \citep[v3.0.5,][]{GALFIT2002,PengHIR10} to perform a decomposition of the 2D surface brightness profile as traced by the old stellar population; toward this goal, we used the PanSTARRS \textit{i}-band image.

To model Source 1 we adopted a compact Sersic profile, which we interpret as a bulge, a truncated Sersic profile for the dumbell-like feature, and a distorted exponential disk for the faint stellar envelope and the irregularities associated with the ``dumbell''. 

The main body of Source 2 was modelled with a Sersic profile and a nuclear point source; the ``tail" was modelled with a Sersic component. An additional Sersic profile was used to model a nearby faint galaxy located northward of the main targets, three other nearby objects were modelled as unresolved sources.

We derived a semi-empirical point spread function (\textsc{psf}) by fitting to a nearby bright star (ra, dec: 08:53:09.30, +16:25:50.2) the model adopted by the PanSTARRS team\footnote{\url{https://outerspace.stsci.edu/display/PANSTARRS/PS1+PSF+photometry+of+detections\#PS1PSFphotometryofdetections-ThePSFModel}}:

\begin{align}\nonumber
f(z) &= A \times (1 + kz + z^{1.67})^{-1},\\ \nonumber 
z &= \left(\frac{x^2}{2\sigma^{2}_{xx}} + \frac{y^2}{2\sigma^{2}_{yy}} + \sigma_{xy}xy\right),
\end{align}

\noindent obtaining $k=0.6$, $\sigma_{xx} = 1.83$, $\sigma_{yy} = 1.82$, $\sigma_{xy}=-0.01$. A model \textsc{psf} was built (setting the amplitude $A=1$) and given in input to \textsc{galfit}. Both \textsc{psf} image and convolution box were 400 $\times$ 400 pixels.

The resulting model and residuals are presented in Fig.\ref{fig: galfit_images}. Subcomponents and best fitting parameters are presented in Appendix \ref{app: galfit}.

% ===========
% Galfit results
% ===========
\begin{figure*}%[th]
\center $
\begin{array}{ccc}
\includegraphics[trim=0cm 0cm 2.5cm 0cm, clip=true, scale=0.5]{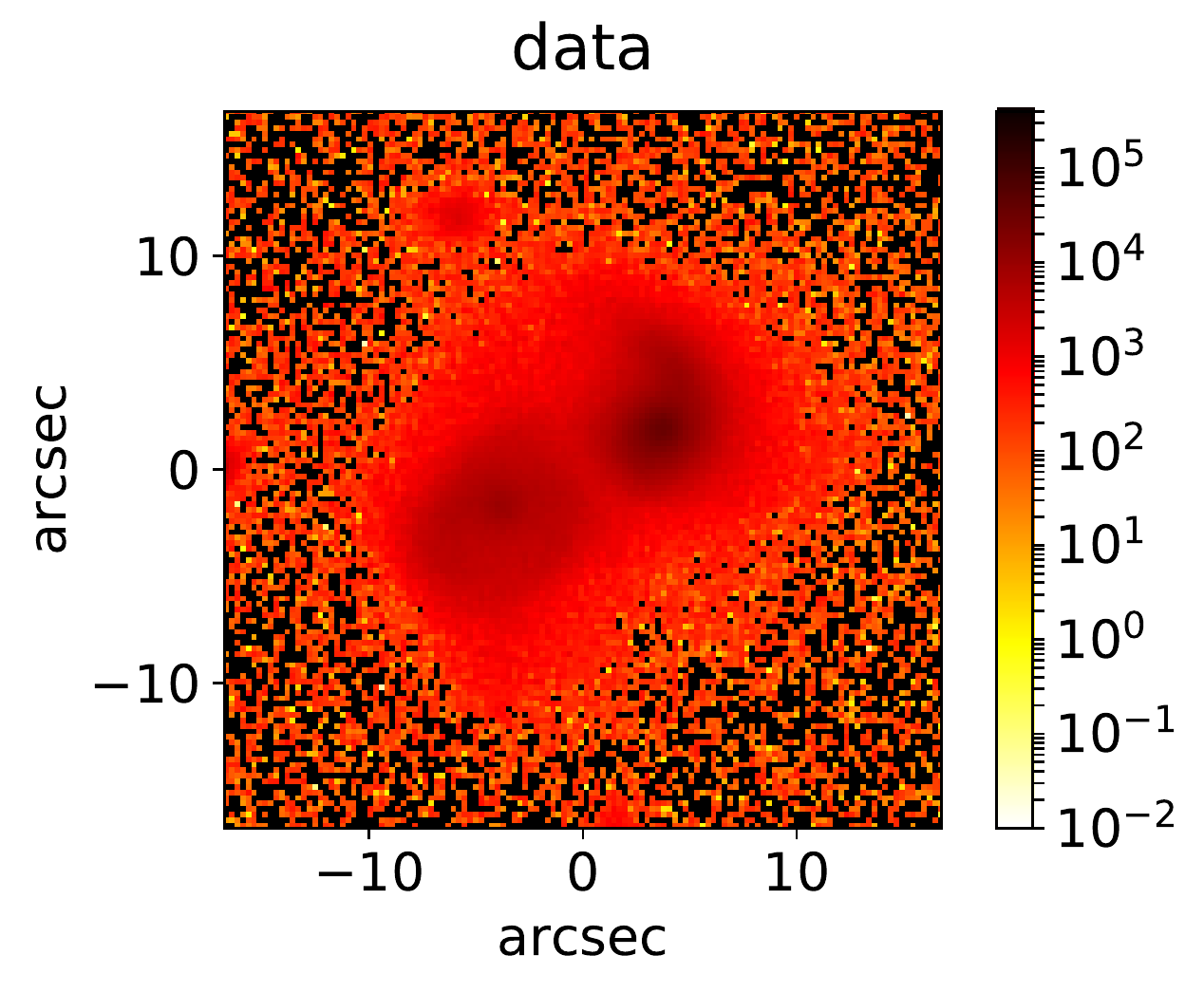} & 
\includegraphics[trim=2.2cm 0cm 0cm 0cm, clip=true, scale=0.5]{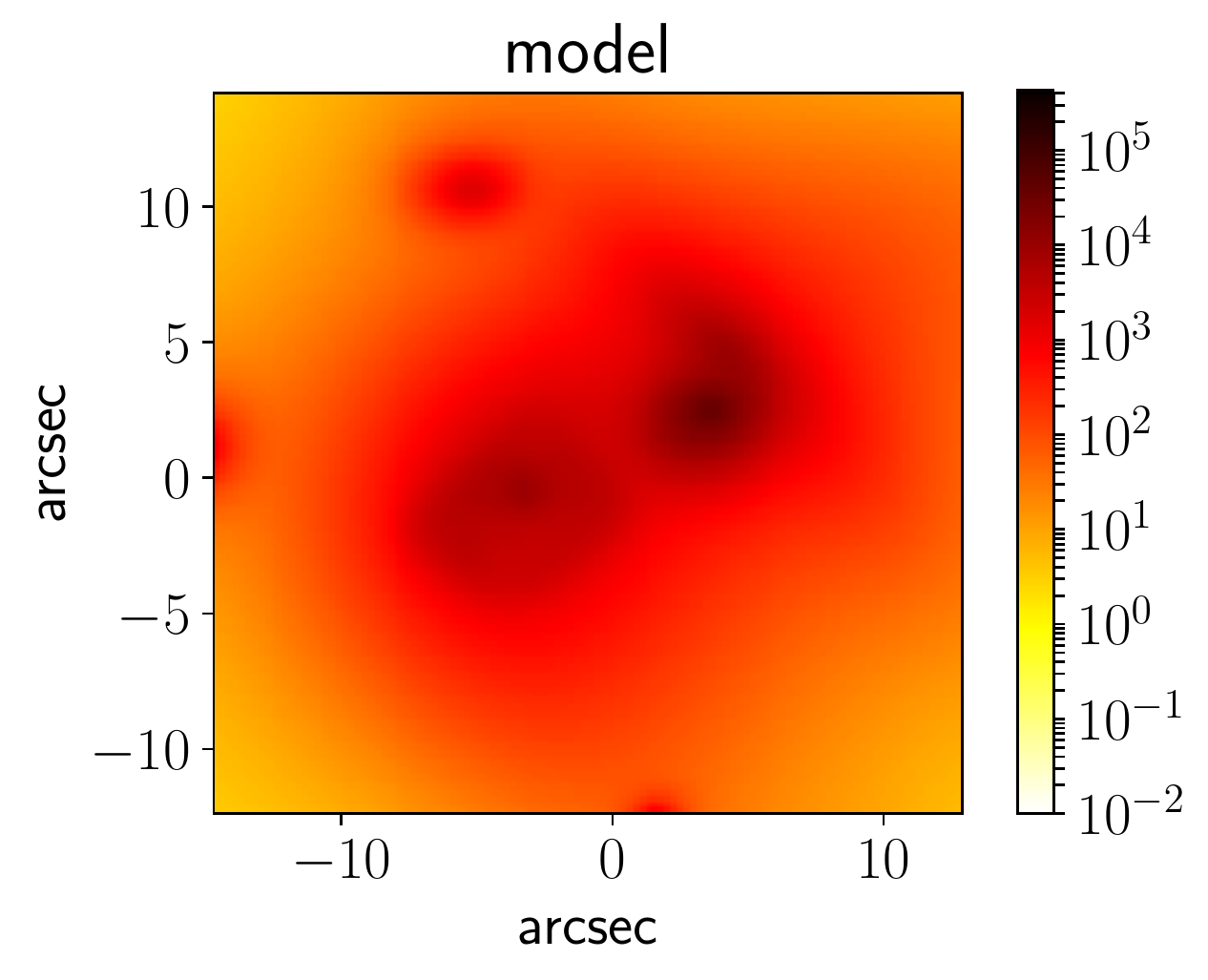} & 
\includegraphics[trim=2.2cm 0cm 0cm 0cm, clip=true, scale=0.5]{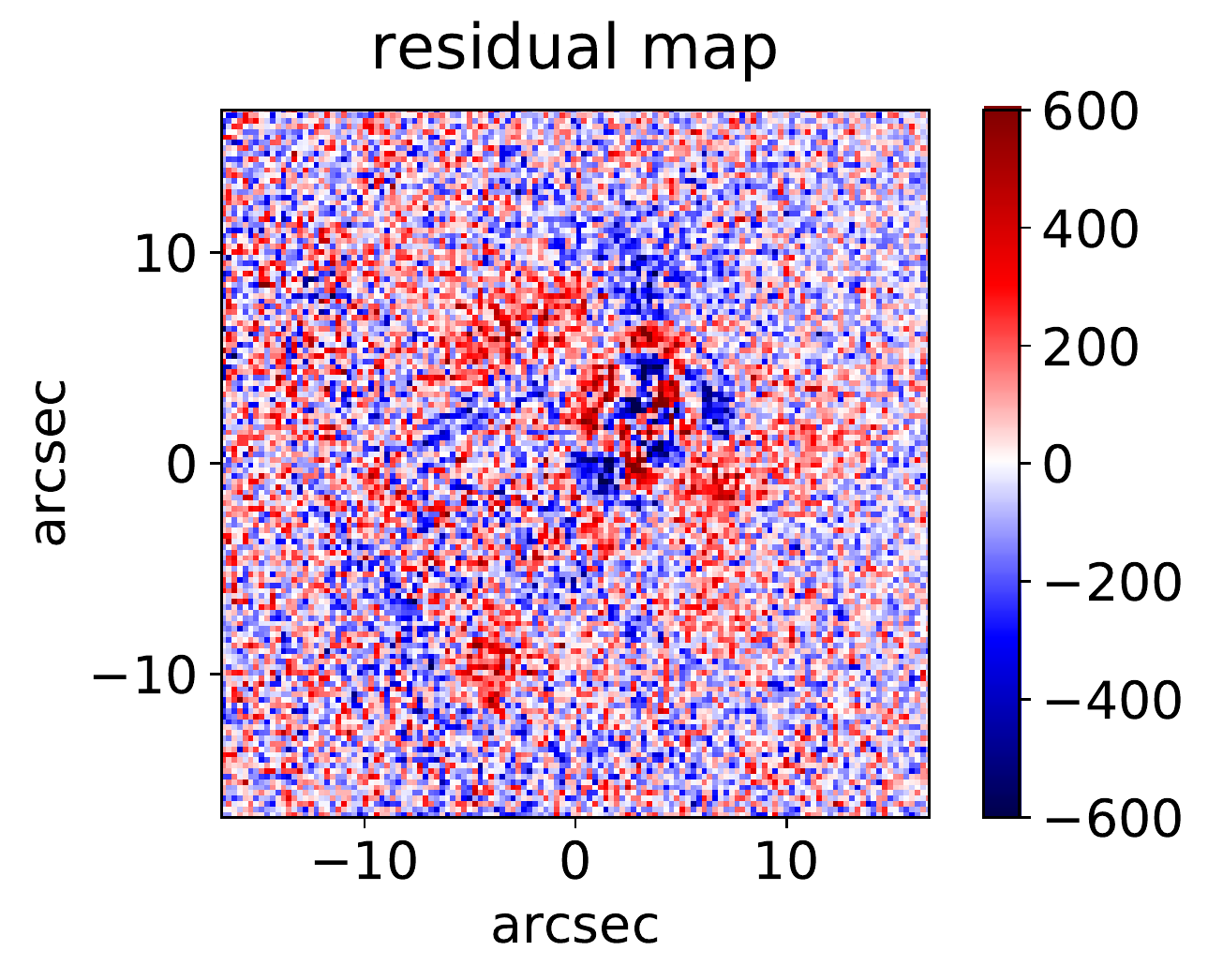}\\
\end{array}$
\caption{Data, model and residual in counts, as obtained by fitting the PanSTARRS \textit{i}-band image with \textsc{galfit}. The same colour bar is adopted for the panels showing the data and the model.}
\label{fig: galfit_images}
\end{figure*}

% --------------------------------------------------------------------------------------------------------------------------------------------
\subsection{Stellar mass ratio}
% --------------------------------------------------------------------------------------------------------------------------------------------
To estimate the total stellar mass ratio we used the $r$ and $i$ mean Kron magnitudes \citep{Kron80} from the PanSTARRS catalog\footnote{\url{http://archive.stsci.edu/panstarrs/search.php}} in conjunction with the scaling relation of \citet{BellMK03}:

\begin{align}\nonumber
log_{10}{\frac{M_{\star}}{M_{\odot}}} = -0.4 (M_{i} - M_{i,\odot}) + 0.006 + 1.114(r-i),
\end{align}

\noindent where M$_{i}$ is the absolute $i$-band magnitude of the galaxy under consideration, M$_{i,\odot} = 4.53$ \citep{BlantonR07} is the absolute AB magnitude of the Sun in the i-band\footnote{The Sun magnitude provided by \citet{BlantonR07} for the \textsc{SDSS} magnitude system was transformed to the PanSTARRS1 system using the linear relation provided by \citet{TonrySL12}, that is $(i_{P1}-i_{SDSS}) = 0.004 - 0.014(g-r)_{SDSS}$, with M$_{i,\odot} = 4.53$, M$_{g,\odot}=5.12$, and M$_{r,\odot}=4.64$. The resulting correction is negligible.}, $r$ and $i$ are the Kron magnitudes of our targets. Using the values in Table \ref{tab: mags}, we obtained log(M$_{1,\star}$/M$_{\odot}$) = 10.7, and log(M$_{2,\star}$/M$_{\odot}$) = 10.9, which correspond to a mass ratio M$_{1}$/M$_{2}$ = 0.7.

We preferred not to use magnitudes estimated via \textsc{galfit} because of a low-luminosity component (component \#4 in Fig.\ref{fig: galfitp}) which fits the extended faint stellar envelope, but also part of the main body of Source 1

\textit{Uncertainties}. Uncertainties on the determination of each mass are dominated by systematics on the distance estimate, however these are expected to have a similar effect on each mass, not on the mass ratio. For the sake of completeness, statistical uncertainties derived via error propagation assuming the uncertainties indicated in Table \ref{tab: mags}, and adopting an uncertainty of 1 Mpc on both distances, are $\Delta$log(M$_{\star}$/M$_{\odot}$) $= 0.03$ for both galaxies, which implies $\Delta M_{1}/M_{2} = 0.05$.

% =========
% Table summary
% =========
\begin{table*}%[t]
\begin{center}
\caption{Best fitting parameters for the optical spectra. (1) Fitted spectral feature; (2) redshift; (3) velocity dispersion; (4) flux; (5) Gauss-Hermite coefficients; (6) component ID. A `d' following the wavelength indicates that the flux refers to the emission line doublet. Velocity dispersions for Source 1 were not corrected for the instrumental broadening because of the uncertain effective resolution of the instrument. Assuming an instrumental resolution $\Delta \lambda = 13.1$ \AA\ for a slit width of 1$^{\prime\prime}$, the instrumental velocity dispersions, at the observed wavelengths are: $\sigma_{\star} = 271$ km s$^{-1}$ (at 6150 \AA), $\sigma_{H\beta} = 322$ km s$^{-1}$, $\sigma_{[OIII]} = 313$ km s$^{-1}$, $\sigma_{[OI]} = 249$ km s$^{-1}$, $\sigma_{H\alpha} = 239$ km s$^{-1}$, $\sigma_{[NII]} = 238$ km s$^{-1}$, $\sigma_{[SII]} = 233$ km s$^{-1}$. Velocity dispersions for Source 2 are already corrected for instrumental broadening.}  
\scalebox{1}{
\begin{tabular}{l llllc}
\hline
feature	 				&  	z					& $\sigma$ (km/s)		& F ($10^{-17}$ erg/s/cm$^2$)	&	h$_{3}$, h$_{4}$	& component ID	\\
(1)						& 	(2)					& (3)					& (4)						& 	(5)				& (6)\\
\hline
& &    \\
\multicolumn{6}{c}{Source 1}\\
\hline
&&\\
stars						& 0.0647 $\pm$ 0.0001		& 284 $\pm$ 38		& -						& -					& \\
H$\beta$					& 0.06470	 $\pm$ 0.00004		& 317 $\pm$ 10		& 225 $\pm$ 8				& -					& 1\\
$[$OIII$]\lambda 5007d$		& 0.06459	$\pm$ 0.00001		& 296 $\pm$ 2			& 1448 $\pm$ 3			& -					& 2\\
$[OI]\lambda 6300d$			& 0.0649 $\pm$ 0.0001		& 275 $\pm$ 24		& 62 $\pm$ 5				& -					& 3\\
H$\alpha$					& 0.06472	 $\pm$ 0.00001		& 259 $\pm$ 2			& 692 $\pm$ 5				& -					& 4\\
$[NII]\lambda 6583d$		& 0.06472	 $\pm$ 0.00001		& 259 $\pm$ 2  		& 416 $\pm$ 5				& -					& 4\\
$[SII]\lambda 6716$			& 0.06472	 $\pm$ 0.00001		& 259 $\pm$ 2			& 152 $\pm$ 4				& -					& 4\\
$[SII]\lambda 6731$ 			& 0.06472	$\pm$ 0.00001		& 259 $\pm$ 2			& 102 $\pm$ 4				& -					& 4\\
&&\\
\multicolumn{6}{c}{Source 2}\\
\hline
& &    \\
stars						&	0.06417 $\pm$ 0.00003	& 164 $\pm$ 9			& -						& -0.058 $\pm$ 0.02, 0.074 $\pm$ 0.017\\
H$\beta$					& 	0.06377 $\pm$ 0.00001	& 132 $\pm$ 3			& $534 \pm 10$				& -					& 1\\
$[$OIII$]\lambda 5007d$		& 	0.06378 $\pm$ 0.00002	& 171 $\pm$ 5 			& $497 \pm 13$				& -					& 2\\
$[OI]\lambda 6300d$			& 	0.06416 $\pm$ 0.00004	& 150 $\pm$ 14		& $147 \pm 10$ 				& -					& 3\\
H$\alpha$					&	0.064325 $\pm$ 0.000003	& 139 $\pm$ 1 			& $2454 \pm 19$ 				& -					& 4\\
$[NII]\lambda 6583d$		& 	0.064325 $\pm$ 0.000003	& 139 $\pm$ 1			& $1657 \pm 16$  			& -					& 4\\
$[SII]\lambda 6716$			&	0.064325 $\pm$ 0.000003	& 139 $\pm$ 1			& $453 \pm 9$ 				& -					& 4\\
$[SII]\lambda 6731$			&	0.064325 $\pm$ 0.000003	& 139  $\pm$ 1			& $364 \pm 7$ 				& -					& 4\\
&&\\
\hline
\end{tabular}}
\label{tab: eml}
\end{center}
\end{table*}

% --------------------------------------------------------------------------------------------------------------------------------------------
\subsection{Star-formation rates}
% --------------------------------------------------------------------------------------------------------------------------------------------
We adopted the relation derived by \citealt{HopkinsMN03} to estimate star-formation rates using the $u$-band magnitudes:

\begin{align}\nonumber
SFR_{u} (M_{\odot} yr^{-1}) = \left[\frac{10^{-0.4(M_{u} - 34.10)}}{1.81 \times 10^{21}} \left(\frac{F_{\mathrm{H}\alpha}/F_{\mathrm{H}\beta}}{2.86}\right)^{2.061}\right]^{1.186}
\end{align} 

\noindent where M$_{u}$ is the absolute magnitude derived using Petrosian magnitudes from SDSS-DR13 (m$_{u,1} = 18.16$ , and m$_{u,2} = 17.58$), corrected for extinction (A$_{u} = 0.07$), and assuming the luminosity distances derived in Sec. \ref{subsec: interact}. F$_{\mathrm{H}\alpha}$ and F$_{\mathrm{H}\beta}$ are the stellar-absorption corrected emission line fluxes from Table \ref{tab: eml}. We obtained \textsc{sfr$_{1}$} $\approx$ 1.5 M$_{\odot}$ yr$^{-1}$ and \textsc{sfr$_{2}$} $\approx$ 7.2 M$_{\odot}$ yr$^{-1}$.

With the estimated \textsc{sfr} we evaluated the contribution of high-mass X-ray binaries to the X-ray luminosity. We used the relation derived by \citealt{GrimmGS03}:

\begin{align}\nonumber
L_{2-10\ \mathrm{keV}} &= 2.6 \times 10^{39} \times \mathrm{SFR}^{1.7}\ [M_{\odot} yr^{-1}], \mathrm{SFR} < 5 M_{\odot} yr^{-1}\\ \nonumber
L_{2-10\ \mathrm{keV}} &= 6.7 \times 10^{39} \times \mathrm{SFR}\ [M_{\odot} yr^{-1}], \mathrm{SFR} > 5 M_{\odot} yr^{-1}
\end{align}

\noindent finding that the hard X-ray luminosity due to X-ray binaries amounts to L$_{2-10\ \mathrm{keV},1} \approx 5_{-3}^{+6} \times 10^{39}$ erg s$^{-1}$, and L$_{2-10\ \mathrm{keV},2} \approx 5_{-2}^{+1} \times 10^{40}$ erg s$^{-1}$.

% ================================================================================
\section{Discussion}  \label{sec_discuss}
For two galaxies to make up a dual-AGN system it is necessary that the galaxies are interacting, and that the two nuclei bear the characteristics of ``active nuclei''. 
We address the presence of these characteristics in the following sections.

% --------------------------------------------------------------------------------------------------------------------------------------------
\subsection{Evidence for interaction}
\label{subsec: interact}
% --------------------------------------------------------------------------------------------------------------------------------------------
Visual inspection of the optical images of the two galaxies (Fig.\ref{fig: xoptical}) shows a prominent elongation departing from Source 2 and bending toward Source 1 - possibly a tidal stream. Faint emission is also present in the surroundings of the two galaxies - possibly stars scattered during the gravitational interaction of the two objects. The morphology of the system, suggests, therefore that the galaxies are indeed interacting.
 
Optical spectra allowed us to estimate their redshift, which we presented in Table \ref{tab: eml}. As the gas emission lines can be affected by non-gravitational motions (i.e. inflows and outflows), biasing the redshift estimate, we derived the redshift from the stellar absorption features obtaining 
$z_{1} = 0.0647 \pm 0.0001$ and $z_{2} = 0.06417 \pm 0.00003$. The difference between the two redshifts corresponds to a velocity difference ($v = cz$) of 160 km s$^{-1}$, a value which can be easily accounted for by the peculiar dynamics of two interacting galaxies.

The picture depicted above supports the hypothesis that the galaxies can be considered an interacting system. 

We used Ned's Wright cosmology calculator\footnote{\url{http://www.astro.ucla.edu/\%7Ewright/CosmoCalc.html}} \citep{Wright2006} to derive the luminosity distance corresponding to the adopted cosmology (Sec. \ref{sec: intro}) and to the redshifts specified above, obtaining the luminosity distances $D_{L,1} \approx 291$ Mpc, and $D_{L,2} \approx 288$ Mpc. Considering that the angular separation between the nuclei is approximately 8$^{\prime\prime}$, or 11 kpc, the physical separation between the two galaxies would be approximately 3 Mpc (about four times the separation between the Milky Way and the Andromeda galaxy), however, as noted above, the difference in the measured redshifts could be due to the peculiar dynamics that two interacting galaxies would experience, therefore we believe it likely that the actual distance between Source 1 and Source 2 is much smaller than 3 Mpc.
\vskip10pt

\textit{Uncertainties.} The uncertainties on the redshift estimate quoted above is the uncertainty in the result of the fit as measured from Monte Carlo simulation (Sec.\ref{subs: fitoptical}). 

The \textsc{rms} scatter in the wavelength calibration of the \textsc{acam} spectrum ($\delta\lambda = 0.5\mathrm{\AA}$, Sec.\ref{sec_result}) corresponds to $\delta z = 0.00008$ at 6500 $\mathrm{\AA}$, or 23 km s$^{-1}$; the uncertainty due to the fitting procedure quoted above is $\delta z = 0.0001$, or $\delta v\approx$ 30 km s$^{-1}$. The resulting statistical uncertainty $(\sqrt{\delta z_{1}^{2} + \delta z_{2}^{2}}$) is $\delta z \approx$ 0.0001, or 38 km s$^{-1}$. 
The estimated systematic uncertainty ($\delta \lambda = 0.1 \mathrm{\AA}$, Sec.\ref{sec_result}) corresponds to 5 km s$^{-1}$. 

The uncertainty on the redshift estimated from the \textsc{sdss} spectrum is dominated by the uncertainty on the wavelength as recovered from the fit, corresponding to 9 km s$^{-1}$.

% --------------------------------------------------------------------------------------------------------------------------------------------
\subsection{Nuclear activity}
\begin{figure*}%[th]
\center $
\begin{array}{ccc}
\includegraphics[trim=0.7cm 0cm 0.3cm 0cm, clip=true, scale=0.5]{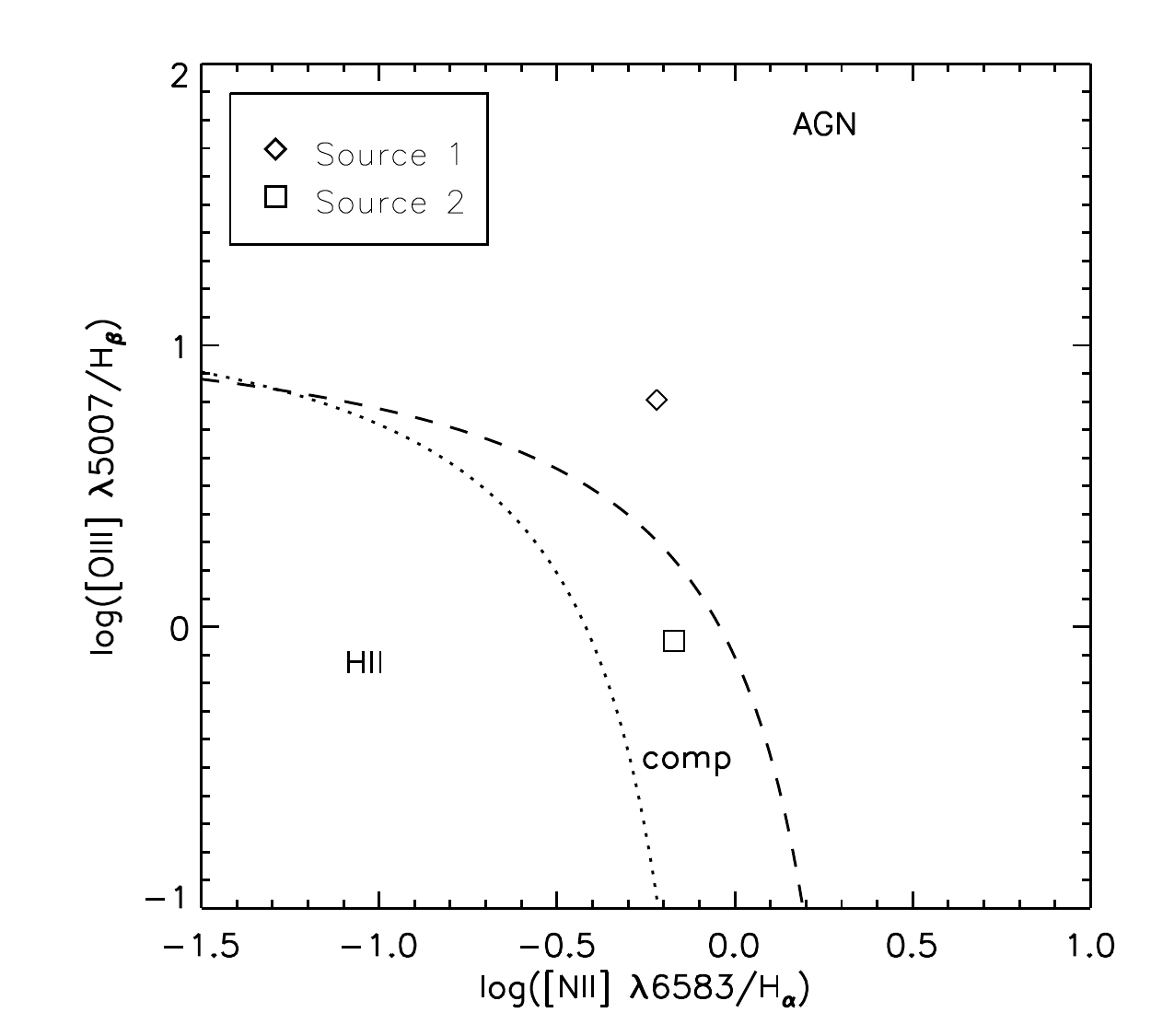}&
\includegraphics[trim=0.7cm 0cm 0.3cm 0cm, clip=true, scale=0.5]{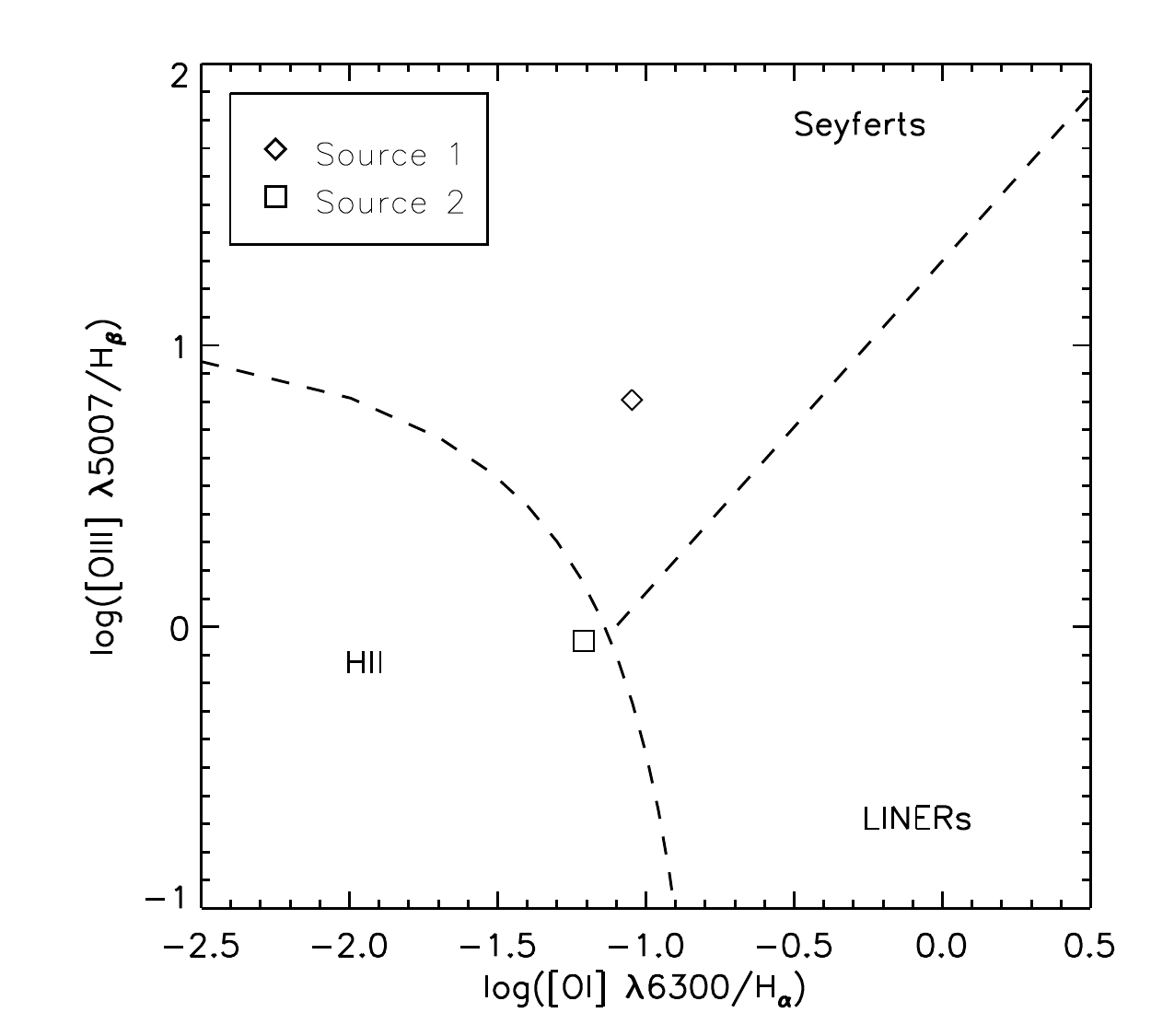}&
\includegraphics[trim=0.7cm 0cm 0.3cm 0cm, clip=true, scale=0.5]{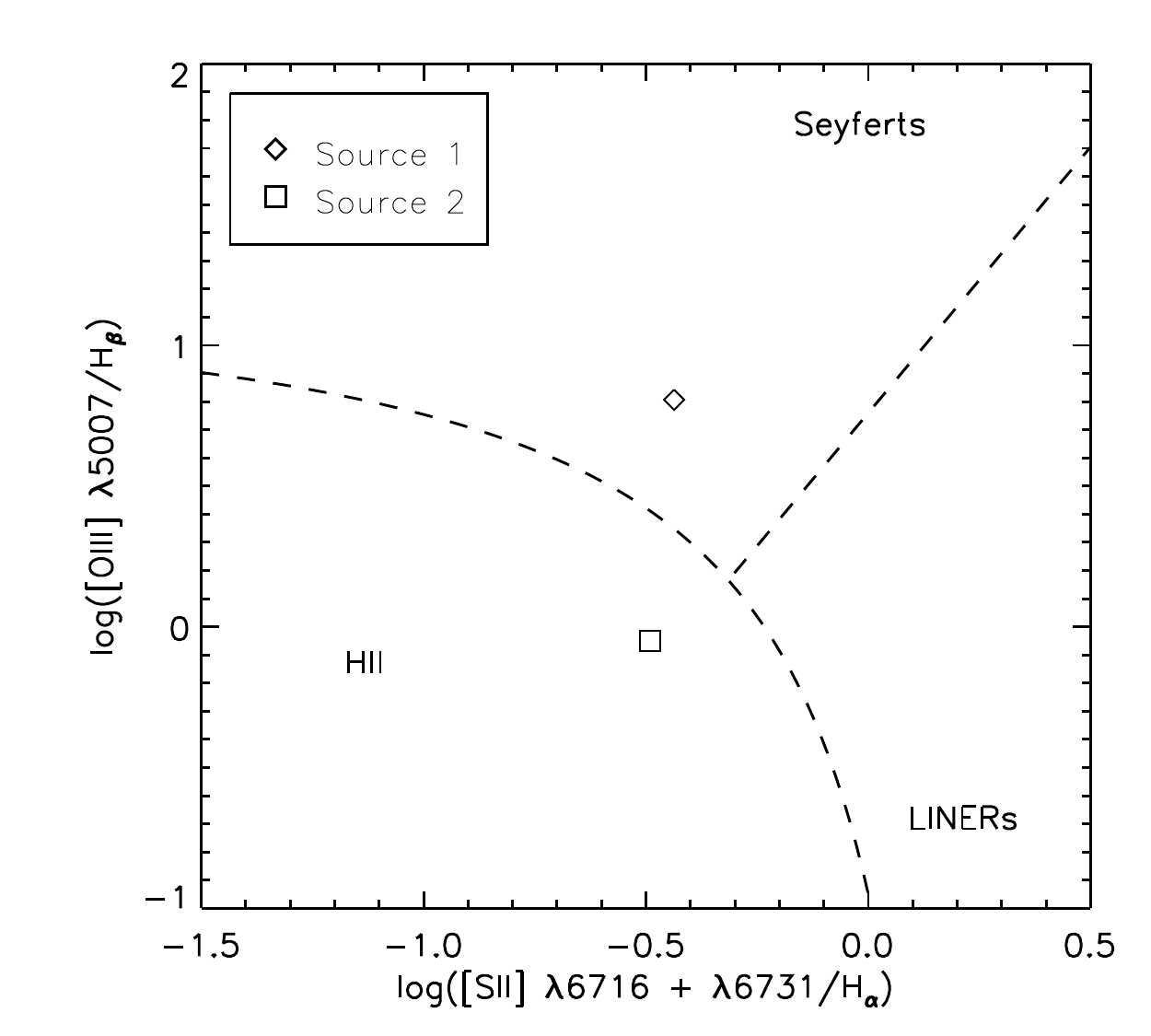}\\
\end{array}$
\caption{Baldwin-Phillips-Terlevich diagrams for the two galaxies. The dotted and the dashed lines in the left panel mark the empirical and the theoretical maximum starburst lines from \citet{KauffmannHT03} and \citet{KewleyEtAl2006}, respectively. Error bars are of the order of the symbol size. Clearly, the ratios of the emission lines found in the spectrum of Source 1 are consistent with it harbouring an active SMBH, whereas the conclusion for Source 2 is ambiguous.}
\label{fig: bpt}
\end{figure*}

% --------------------------------------------------------------------------------------------------------------------------------------------
\textit{Optical.} Emission line ratios can be used as diagnostics to infer the ionization mechanism. This is of particular interest in our case, as we wish to understand whether AGNs reside in the nuclei of the two galaxies. 
We plotted the relevant line ratios in Baldwin-Phillips-Terlevich diagrams \citep[\textsc{bpt},][]{BPT81} adopting the boundaries proposed by \citet{KewleyEtAl2006}. Results are shown in Fig.\ref{fig: bpt}.

The dominant ionisation mechanism in Source 1 is clearly AGN activity. Instead, star formation is the main ionisation mechanism in Source 2, which also shows a more uniform blue colour. However, it is worth noting that the [OI]/H$\alpha$ ratio is close to the boundary between HII and AGN ionisation, and the [NII]/H$\alpha$ line ratio falls in  the ``composite" region, that is in between the empirical and theoretical maximum starburst lines. It is, therefore, possible that a weak AGN resides in Source 2 as well.
\vskip10pt

\textit{Mid-infrared.} As shown in Sec.\ref{subsec: col}, mid-infrared colours derived from \textsc{wise} photometry \citep{WrightEM10} are characteristic of AGNs for Source 1. The color of Source 2, instead, falls in between the locus of Starbursts and \textsc{lirg}s.

When interpreting this result the reader should note that the \textsc{wise psf} has a large \textsc{fwhm} (between 6$^{\prime\prime}$ and 7$^{\prime\prime}$). The colours are, therefore, representative of the whole host galaxies, and the contribution of an hypothetical weak \textsc{agn} residing in Source 2 could easily be overwhelmed by the stellar emission.
\vskip10pt

\textit{X-ray.} Hard X-ray emission (i.e. at energies in excess of 5 keV) is an excellent tracer of AGN activity. Such emission is indeed associated with Source 1, which also shows optical line-ratios and mid-infrared colours typical of AGN ionisation. From the fit of the X-ray spectrum we derived $L_{x,2-10 keV} \approx 5 \times 10^{42}$ erg s$^{-1}$, and from the SFR we estimated that the X-ray contribution from X-ray binaries is two-to-three orders of magnitudes lower than the observed values. These numbers firmly identify Source 1 as an active galaxy.

On the other hand, we detected no emission above 2 keV in Source 2. The luminosity extrapolated from the fit in the energy range 2 - 10 keV is $L_{x} = 1.7_{-1.2}^{+1.3} \times 10^{40}$ erg s$^{-1}$; this is consistent with the value expected from the observed SFR ($5_{-2}^{+1} \times 10^{40}$ erg s$^{-1}$), however, the predicted X-ray luminosity derived using the scaling relation of \citealt{GrimmGS03} refers to the luminosity of the galaxy as a whole; instead, the X-ray luminosity quoted for Source 2 was derived from a region including no more than about 50\% of the galaxy mass (this is approximately the mass ratio between the boxy-bulge and the tadpole body highlighted in the sdss $u$-band image, Fig.\ref{fig: xoptical}). It might be an indication that a weak AGN is present in Source 2. 

To frame these results in a broader context, the reader should recall that X-ray emission due to star formation is known to be as high as a few $10^{41}$ erg s$^{-1}$ \citep[e.g.][]{ZezasAHW01}; ultra-luminous X-ray source have typical luminosities in the range $10^{39} - 10^{40}$ erg s$^{-1}$; and hyper-luminous X-ray source are known to reach $10^{42}$ erg s$^{-1}$ \citep[e.g.][]{FarrellWB09,HeidaJT15}.

As in \citealt{LiuCS13} and \citealt{ShangguanLH16} we find hard X-ray emission in the nucleus displaying redder colours.
\vskip10pt

To summarise, while the active nature of Source 1 can be established beyond reasonable doubt, there is only tentative evidence for an AGN in Source 2.

% --------------------------------------------------------------------------------------------------------------------------------------------
\subsubsection{A note on nuclear point sources}\label{subsec: galfit_note}
% --------------------------------------------------------------------------------------------------------------------------------------------
As stated in Section \ref{subsec: galfit}, a nuclear point source (NPS) was used to fit the optical 2D brightness profile of Source 2, while a model without a NPS was adopted for Source 1. The latter can also be modelled with the inclusion of a faint nuclear point source of magnitude $i = 20.6$, however the model does not show any significant improvement, as the reduced chi square remains unaltered, and residuals show very little difference. The parameters for the resulting nuclear Sersic component also show minor changes, and are mostly consistent with the parameters obtained without a NPS. For these reasons we adopt the simplest model, which does not include a NPS.

The absence of a bright NPS in Source 1 can be explained noting that Source 1 is host to a Type 2 AGN (lacking permitted broad emission lines in the optical spectrum); therefore, according to the unified model of AGNs \citep[e.g][]{UrryP95}, a direct view of the SMBH accretion disk (physical counterpart for the nuclear point source) is prevented because of dust obscuration - which we estimated to be relatively high in Source 1 (n$_{H} \approx 5 \times 22$ cm$^{-2}$, Sec.\ref{subsec: xrayf}).

On the other hand, the NPS in Source 2, a galaxy dominated by star-formation and characterised by a lower nuclear obscuration (n$_{H}\approx 2 \times 10^{20}$ cm$^{-2}$), could be ascribed to a nuclear star cluster.

% =================

% --------------------------------------------------------------------------------------------------------------------------------------------
\subsection{X-ray vs [OIII] luminosity}\label{sebsec: xo3}
% --------------------------------------------------------------------------------------------------------------------------------------------
The flux of the [OIII]$\lambda$5007 emission line originates at scales of hundreds to thousands of parsecs from the ionising AGN, and it is therefore free of the effects of the nuclear absorption acting on the X-ray emission; extinction due to the large-scale dust distribution can be inferred from the Balmer decrement. As a result, the [OIII] luminosity has been proposed as a proxy for the AGN true luminosity, and it has been used to infer the X-ray absorption \citep[e.g.][]{BassaniDM99}. 

\citet[][]{LiuCS13} showed tentative evidence that optically-selected kpc-scale dual AGNs do not follow the trend defined by isolated AGNs, displaying a lower X-ray-to-[OIII] ratio; perhaps an indication of the large amount of gas and dust funnelled toward to the center during the merger process. 

We derived the extinction-corrected [OIII] luminosity as:

\begin{align}\nonumber
L_{[OIII]}^{c} = L_{[OIII]} \left(\frac{F_{H\alpha}/F_{H\beta}}{3}\right)^{2.94},
\end{align}

\noindent where L$_{[OIII]}$ is the observed luminosity of the [OIII] emission line as derived from the fluxes in Table \ref{tab: eml} (respectively L$_{[OIII],1}$ = 1.46 $\times$ 10$^{41}$ erg s$^{-1}$ and L$_{[OIII],2}$ = 4.94 $\times$ 10$^{40}$ erg s$^{-1}$), F$_{H\alpha}$ and F$_{H\beta}$ are the observed fluxes of the H$\alpha$ and H$\beta$ emission lines, and an intrinsic Balmer decrement of 3.0 was assumed \citep{OsterbrockF06_book}. This resulted into L$_{[OIII],1}^{c} \approx 2 \times 10^{41}$ erg s$^{-1}$, and L$_{[OIII],2}^{c} \approx 2 \times 10^{39}$ erg s$^{-1}$.

The location of our targets in a L$_{x,2-10keV}$ vs L$_{[OIII]}$ is shown in Fig.\ref{fig: xvso3}, along with a comparison sample of Seyferts and \textsc{liner}s drawn from the Palomar survey of nearby galaxies \citep{HoFS95} by \citet{PanessaBC06}: the two nuclei fall within the locus defined by single AGNs. With the caveat that the L$_{x,2-10keV}$ for Source 2 is an extrapolation based on the emission at energies below 2 keV, the result might be an indication that Source 2 does harbour a low-luminosity AGN.

% ======================================
\subsection{Eddington ratio}
% ======================================

To estimate the Eddington ratio of Source 1 (the galaxy with strong evidence for AGN activity), $\lambda_{E}  = L_{bol}/L_{Edd}$, we need to estimate the bolometric luminosity (L$_{bol}$) and the Eddington luminosity (L$_{Edd}$).

We computed the bolometric luminosity using scaling relations with the optical and X-ray data. In the optical we adopted the bolometric correction (C$_{[OIII]}$ = L$_{bol}$/$L_{[OIII]}^{c}$) derived by \citet{LamastraBM09} from extinction-corrected [OIII] emission lines. In Sec.\ref{sebsec: xo3} we derived L$_{[OIII],1}^{c} \approx 2 \times 10^{41}$ erg s$^{-1}$, 
which requires a bolometric correction C$_{[OIII]}$ = 87, producing $L_{bol,1} \approx 1 \times 10^{43}$ erg s$^{-1}$.

We also estimated the bolometric luminosity using the scaling relation of \citealt{Ho09}: C$_{X}$ = L$_{bol}$/L$_{X,2-10 keV}\approx 8$, which produces $L_{bol,1} \approx 2 \times 10^{43}$ erg s$^{-1}$. However, \citeauthor{Ho09} argued that a larger data set supports C$_{X} \approx 15.8$. Given this uncertainty, and the consistency between the different estimates, we adopted the value derived from the [OIII] emission line.

The Eddington luminosity is defined as:

\begin{eqnarray}\nonumber
L_{E} = 4 \pi c G M_{\bullet}m_{p} / \sigma_{T},
\end{eqnarray}

\noindent where $c$ is the speed of light, $m_{p}$ the mass of the proton, $\sigma_{T}$ the Thomson cross section, and $M_{\bullet}$ is the mass of the black hole which is responsible for the nuclear activity. 

The mass of the SMBH is very uncertain, nevertheless we present an estimate based on two approaches: the scaling relation between SMBH mass and bulge Sersic index \citep{SavorgnanGM13}, and the scaling relation between SMBH mass and bulge velocity dispersion \citep{FerrareseM2000,GBBetAl00}.

To warrant the use of such scaling relations, one should find evidence for a bulge. The task is not trivial, and multiple lines of evidence should be put forward (such as photometric, possibly from high-resolution images, and kinematical, e.g. \citealt{Gadotti12}). Our claim for the presence of a bulge-like component in Source 1 is based on the \textsc{2d} decomposition of the $i$-band image presented in Section \ref{subsec: galfit}, where we showed that a satisfactory fit can be achieved with the inclusion of a compact Sersic component with low index (\#1 in Fig.\ref{fig: galfitp}), possibly a rotationally-supported bulge.
\vskip10pt

% ======================================
\textbf{M$_{\bullet}$ - Sersic index}: For Source 1 we measured the Sersic index $n_{1} = 0.3 \pm 0.1$. Using the scaling relation presented in \citet{SavorgnanGM13}:

\begin{align}\nonumber
log(M_{\bullet}) = (7.85 \pm 0.14) + (3.38 \pm 1.16)log(n/3),
\end{align} 

\noindent we obtained M$_{\bullet} \sim 10^{4}$ M$_{\odot}$, which corresponds to the Eddington luminosity $L_{E} \sim 10^{42}$ erg s$^{-1}$. The resulting Eddington ratio is $\lambda_{E} \approx 4$.

\textit{Caveats and uncertainties.} The $M_{\bullet} - n$ relation was derived using Sersic indices $n \geq 1$; the correlation, with a scatter of $0.44$ dex, is therefore unconstrained for $n < 1$. 

The structural parameters derived in a multi-component fit could be affected by degeneracies (see the detailed discussion in \citealt{PengHIR10}), and the Sersic index is also affected by orientation and extinction \citep{PastravPT13}. 

Removing or adding azimuthal shape functions (i.e. Fourier modes and bending modes) in the components that we adopted in our \textsc{galfit} model caused variations in the Sersic index which were never larger than 9$\%$. We observed similar variations when varying the initial guess. However, the removal of a low-luminosity component, such as the exponential disk used to model the irregularities of Source 1 and the stellar envelope, caused a variation as large as 20$\%$. This is the value that we assumed to represent the systematic errors in the Sersic index, which corresponds to a systematic uncertainty on the Eddington ratio $\lambda_{E,1} \approx 4_{-2}^{+5}$.

Systematics in the Sersic index due, separately, to the orientation and the dust, as derived from bulge-disk (\textsc{b/d}) decomposition of model galaxies, are always in the range $-0.1 < n^{B/D} - n^{single} < +0.3$, with the correction being between $-0.05$ and $+0.05$ for intermediate inclinations, where $n^{B/D}$ and $n^{single}$ are the Sersic indices of decomposed and single bulges respectively \citep{PastravPT13}. These uncertainties, at intermediate inclinations, correspond to a systematic error in $\lambda_{E,1}$ of $-2, +3$ due to the effect of dust, and $-2, +3$ due to effect of orientation.

% =========
% Table summary
% =========
\begin{table}[t]
\begin{center}
\caption{Stellar masses. (1) Source ID; (2) distance in Mpc, as inferred in Sec.\ref{subsec: interact}; (3) absolute $i$-band magnitude; (4, 5) $r$ and $i$-band magnitude.}  
\scalebox{0.9}{
\begin{tabular}{l llllc}
\hline
Source	&  	D	& M$_{i}$		& r					& i				& log10(M$_{\star}$/M$_{\odot}$)\\
	 (1)	&	(2)	& (3)			& (4)					& (5) 			& (6)\\
\hline
& &    \\
1		& 	291	& -21.2		& 16.49 $\pm$ 0.02		& 16.11 $\pm$ 0.01	& 10.7\\
2		& 	288	& -21.6		&  16.05 $\pm$	0.02		& 15.68 $\pm$ 0.01	& 10.9\\
	
& &    \\
\hline
\end{tabular}}
\label{tab: mags}
\end{center}
\end{table}

% ======================================
\textbf{M$_{\bullet}$ - stellar velocity dispersion}: The observed stellar velocity dispersion for Source 1 is $\sigma_{obs} = 284 \pm 38$ km s$^{-1}$. Given that the stellar velocity dispersion is derived from spectral features distributed over the full spectral range (5000 - 7500 \AA), we corrected this value for the instrumental resolution $\sigma_{ins} = 271$ km s$^{-1}$, that is the theoretical instrumental velocity dispersion at 6150 \AA, assuming the nominal dispersion of 13.1 \AA\footnote{$\sigma_{ins} = \frac{c}{2.355}\frac{\Delta\lambda}{\lambda}$, with $c$ the speed of light, and $\Delta\lambda=13.1$\AA\ the instrumental resolution. The high value of $\sigma_{ins}$, that is the  low resolution of \textsc{acam}, is responsible for the large uncertainty on the ``intrinsic'' stellar velocity dispersion which appears later in the text.}. The resulting intrinsic stellar velocity dispersion ($\sqrt{\sigma_{obs}^2 - \sigma_{ins}^{2}}$) is $\sigma = 84^{+127}_{-84}$ km s$^{-1}$, where we estimated the uncertainty using the error propagation formula for independent variables, assuming no uncertainty on $\sigma_{ins}$. This value can be used in conjunction with the M$_{\bullet} - \sigma$ relation presented in \citet{FerrareseFordRev05}:

\begin{align}\nonumber
\frac{M_{\bullet}}{10^{8}M_{\odot}}  & = (1.66 \pm 0.24)\left(\frac{\sigma}{200\ \mathrm{km s}^{-1}}\right)^{4.86\pm0.43}.
\end{align}

\noindent As in \citet{FerrareseM2000}, we applied an aperture correction to the velocity dispersion normalising it to an aperture of radius 1/8 of the bulge effective radius. Toward this end, we used the prescription of \citet{JorgensenFK95}:

\begin{align}\nonumber
\sigma_{e8} = \sigma_{ap} \left(\frac{r_{e}/8}{r_{ap}}\right)^{-0.04}
\end{align}

\noindent where r$_{e} = 0\farcs6$ is the bulge effective radius as derived from \textsc{galfit} (taking into account the PanSTARRS pixel size of $0.258$ arcsec), $r_{ap} = 1\farcs75$ is the radius of the aperture used in IRAF to extract the spectrum, and $\sigma_{ap} = 84$ km s$^{-1}$ is the observed stellar velocity dispersion corrected for instrumental broadening. We obtained $\sigma_{8} = 95^{+144}_{-95}$ km s$^{-1}$. 

Given the large uncertainty on the stellar velocity dispersion, we estimated an upper limit to the SMBH mass: adopting $\sigma \leq 239$ km s$^{-1}$, we obtained M$_{\bullet} \leq 4 \times 10^{8}$ M$_{\odot}$, and L$_{E} \leq 5 \times 10^{46}$ erg s$^{-1}$. The resulting Eddington ratio is $\lambda_{E} \geq 10^{-4}$.

\begin{figure*}
\begin{center}$
\begin{array}{cc}
\includegraphics[trim=0cm 0cm 0cm 0cm, clip=true, width=0.45\textwidth]{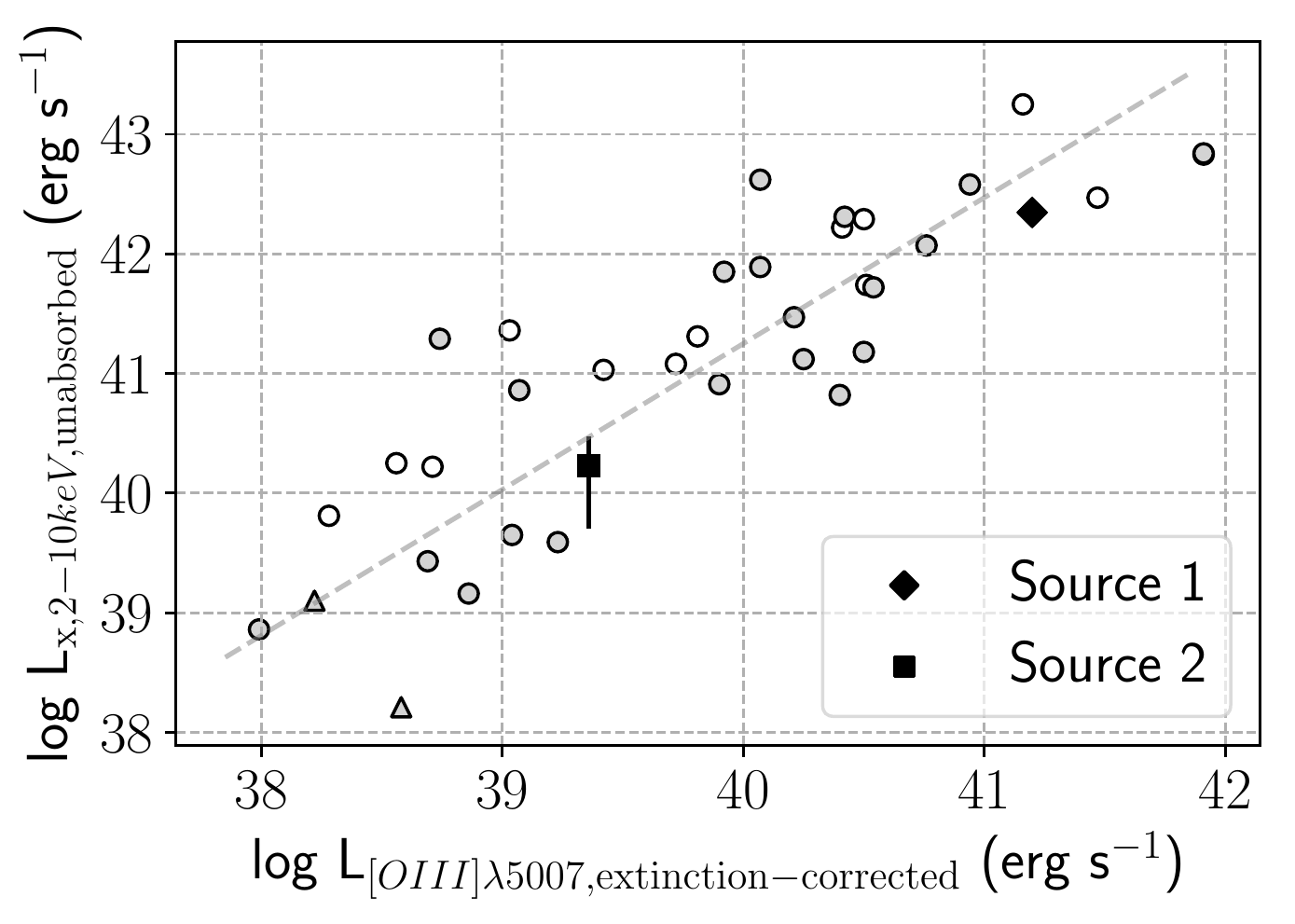} &
\includegraphics[trim=0cm 0cm 0cm 0cm, clip=true, scale=0.57]{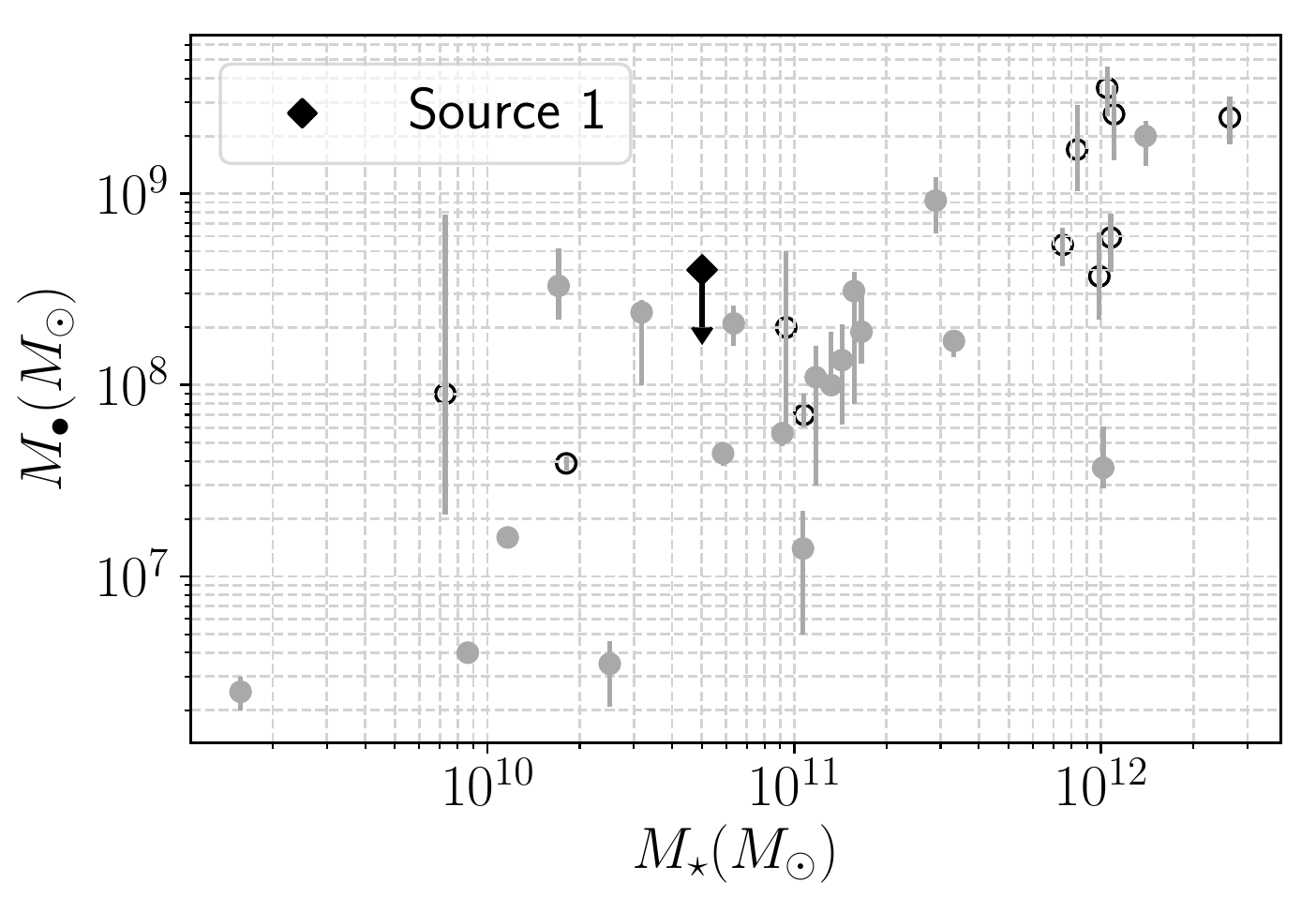}
\end{array}$
\end{center}
\caption{\textit{Left:} Unabsorbed 2 - 10 keV luminosity vs extinction-corrected [OIII] luminosity. The comparison sample is taken from \citet{PanessaBC06} and it includes galaxies classified as Seyfert 1 - 1.5 (open circles), Seyferts 1.8 - 2 (grey circles), and two galaxies classified as \textsc{liner}2/Sy2 (grey triangles). Galaxies for which the classification was marked as uncertain or highly uncertain were not included into this plot. The dashed line is the correlation derived by \citeauthor{PanessaBC06} for the mixed sample of Seyferts (i.e. Log$_{x}$ = (1.22$\pm$0.11) $\times$ Log$_{[OIII]}$ - (7.55$\pm$4.33)). When omitted, errorbars for Source 1 and 2 are smaller than the symbol size. Note that the X-ray luminosity for Source 2 is an extrapolation from the fit. \textit{Right:} black-hole mass $vs$ stellar galaxy mass. The comparison sample consists of galaxies from Table II in \citet[][excluding galaxies for which dynamical models are deemed to be in error]{FerrareseFordRev05}. BH masses are taken from \citet[][see references therein]{FerrareseFordRev05}, virial stellar masses have been estimated as $\alpha$R$_{e} \sigma^{2}$/G, with $\alpha = 5$ \citep{CappellariBB06}, R$_{e}$, the effective radius, and $\sigma$, the stellar velocity dispersion, from \citep{MarconiH2003}. Grey dots: non-active galaxies; black circles: active galaxies (as classified on the N\textsc{asa/ipac} E\textsc{xtragalactic} D\textsc{atabase}).}
\label{fig: xvso3}
\end{figure*}

When compared to other BHs residing in galaxies of similar masses, we can only conclude that the BH in Source 1 is not obviously over-massive, Fig.\ref{fig: xvso3}.
\vskip10pt

To summarise, because of the large uncertainties on the SMBH mass, the Eddington ratio is poorly constrained, with $10^{-4} \leq \lambda_{E} \leq 9$. 

% ======================================

% ================================================================================
\section{Summary and Conclusions} \label{sec_conclusions}
In \textsc{sdss-dr7} we selected galaxies separated by no more than 10 times the sum of their Petrosian radii. Subsequently, we cross-correlated the sample with the XMM serendipitous source catalog (2XMMi) looking for galaxy pairs associated with a single X-ray source (possibly a recoiling SMBH). The search produced two candidates: the hyperluminous X-ray source described in \citet{JonkerTF10}, and the system studied in this paper: a pair of interacting galaxies with stellar mass ratio $q\approx 0.7$, and an X-ray source approximately coincident with the optical nucleus of the Eastern galaxy (Source 1). To further investigate the nature of the system we performed \co observations, which revealed a nuclear X-ray source in the Western galaxy as well (Source 2), and we used \textsc{acam} on the William Herschel Telescope to obtain a new optical spectrum for Source 1.

Source 1 shows hard X-ray emission, low star-formation rate (2 M$_{\odot}$ yr$^{-1}$), a redder nucleus with mid-infrared colours located within the locus of AGN candidates, and optical emission line ratios typical of AGN photoionisation. The active nature of this galaxy is evident.

Source 2, instead, has a softer spectrum with no emission above 2 keV; it has a higher star-formation rate (7 M$_{\odot}$ yr$^{-1}$), indeed mid-infrared colours are typical of Starburst/\textsc{liner}s; the optical emission line ratios are also characteristic of HII regions, however they are close to the demarcation between starburst- and AGN-driven ionisation, with the [NII]/H$\alpha$ emission line ratio falling in the ``composite'' region. 

When plotted on a L$_{x,2-10 keV}$ vs L$_{[OIII]}$ diagram, both nuclei fall within the locus defined by a mixed sample of local AGNs, mostly Seyfert galaxies. Moreover, the 2-10 keV luminosity extracted from the inner region of Source 2 is consistent with the value expected for the galaxy as a whole on the basis of the star-formation rate. With the caveat that the hard X-ray luminosity for Source 2 is an extrapolation based on the soft X-ray spectrum, these are, perhaps, hints that a low-luminosity AGN is buried within Source 2.

Using the relative intensity of the [SII] emission lines, we estimated the electron densities finding that Source 1 has a density $n_{e} < 27$ e$^{-}$ cm$^{-3}$, while Source 2 has an electron density of approximately 200 e$^{-}$ cm$^{-3}$.

We used \textsc{galfit} to perform a 2D decomposition of the light profiles. Both galaxies can be modelled with a inner Sersic component with an index $n < 1$, suggesting the presence of rotationally supported bulges. For Source 1 we estimated the Eddington ratio, $\lambda_{E}$, however the black hole mass, estimated using two scaling relations (M$_{\bullet}$ vs Sersic index, and M$_{\bullet}$ vs stellar velocity dispersion), is highly uncertain leaving $\lambda_{E}$ essentially unconstrained.

While the active nature of Source 1 can be established with confidence, our analysis does not allow a definitive statement about the presence of an AGN within Source 2. However, hints of its existence are tantalising. Bearing in mind the importance of defining a well-characterised, secure sample of AGNs within interacting galaxies, we will further probe the nature of Source 2 with dual-frequency radio observations. The detection of a compact steep-spectrum radio source would definitely prove the active nature of its nucleus.

%%%%%%%%%%%%%%%%%%%%

\section*{Acknowledgements}

The authors thank K. M. L\'{o}pez for carrying out the \textsc{acam} observations, and the anonymous referee for valuable comments and suggestions. DL and PGJ acknowledge support from the European Research Council (ERC) under grant 647208 (PI Jonker). DL thanks the organisers and the participants of the Lorentz Center workshop ``The quest for multiple supermassive black holes: a multi-messenger view'' for constructive and informative discussions; DL also thanks D. Merritt for constructive comments., C. Y. Peng and M. Cappellari for timely and clear assistance with the use of \textsc{galfit} and pPxf, respectively. MAPT acknowledges support via a Ramón y Cajal Fellowship (RYC-2015-17854).

Based on observations made with the William Herschel Telescope, operated on the island of La Palma by the Isaac Newton Group of Telescopes in the Spanish Observatorio del Roque de los Muchachos of the Instituto de Astrofísica de Canarias.
This research has made use of the NASA/IPAC Extragalactic Database (NED), which is operated by the Jet Propulsion Laboratory, California Institute of Technology, under contract with the National Aeronautics and Space Administration.

%%%%%%%%%%%%%%%%%%%%%%%%%%%%%%%%%%%%%%%%%%%%%%%%%%



\bibliographystyle{mnras}
\bibliography{biblio.bib}

%%%%%%%%%%%%%%%%%%%%%%%%%%%%%%%%%%%%%%%%%%%%%%%%%%

%%%%%%%%%%%%%%%%% APPENDICES %%%%%%%%%%%%%%%%%%%%%

\appendix

\section{GALFIT components}\label{app: galfit}
\begin{figure*}%[th]
\center $
\begin{array}{ccc}
\includegraphics[trim=0cm 0cm 0cm 0cm, clip=true, scale=0.45]{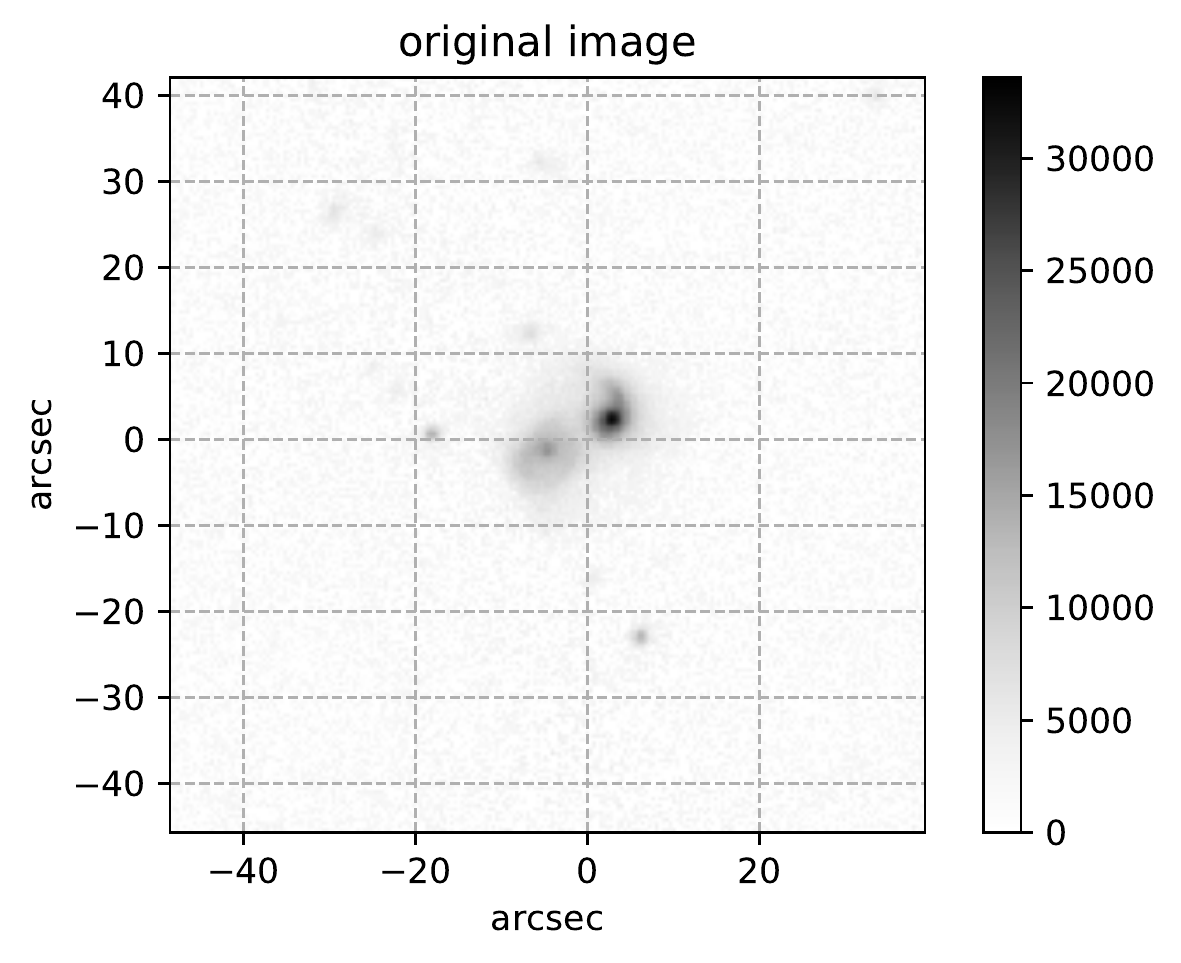} & 
\includegraphics[trim=0cm 0cm 0cm 0cm, clip=true, scale=0.45]{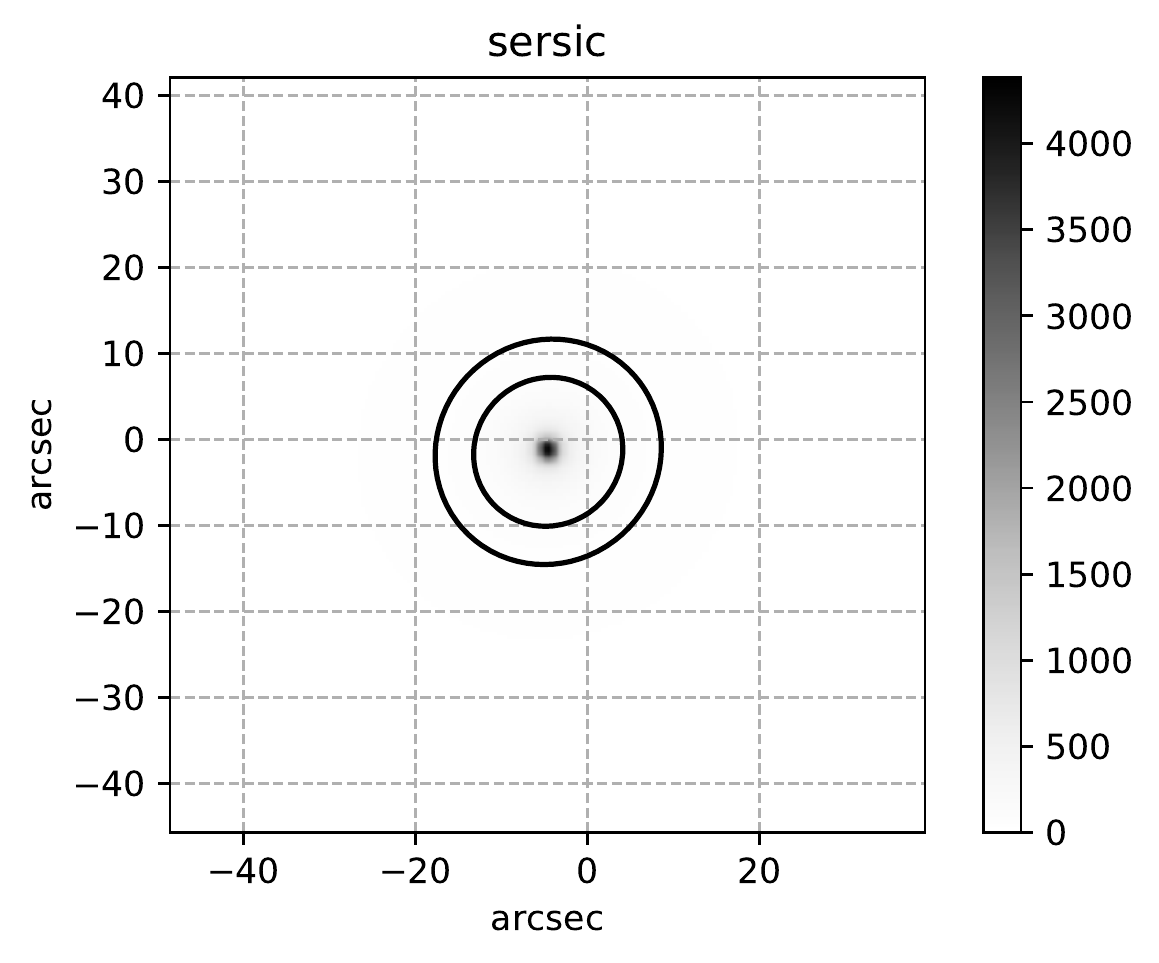} & 
\includegraphics[trim=0cm 0cm 0cm 0cm, clip=true, scale=0.45]{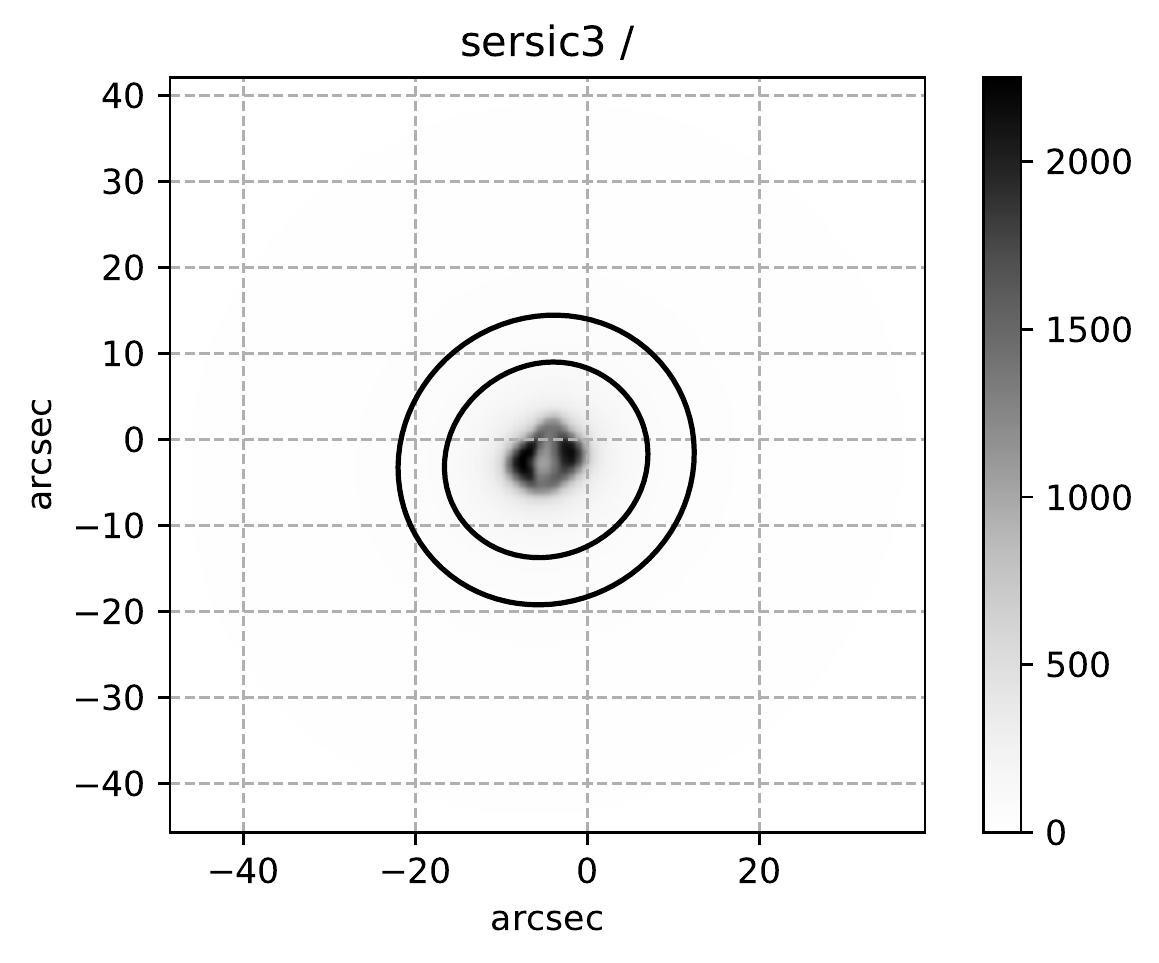}\\
\includegraphics[trim=0cm 0cm 0cm 0cm, clip=true, scale=0.45]{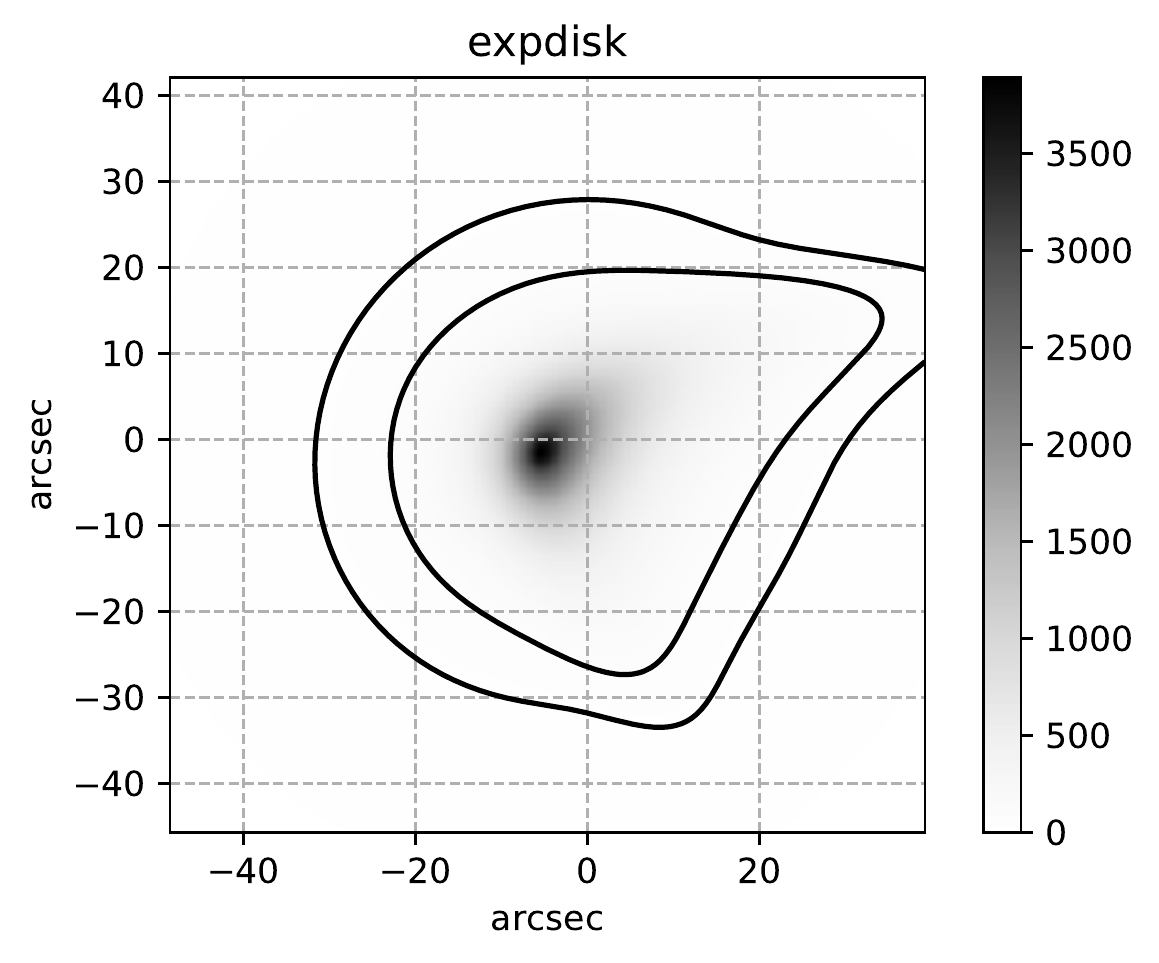} & 
\includegraphics[trim=0cm 0cm 0cm 0cm, clip=true, scale=0.45]{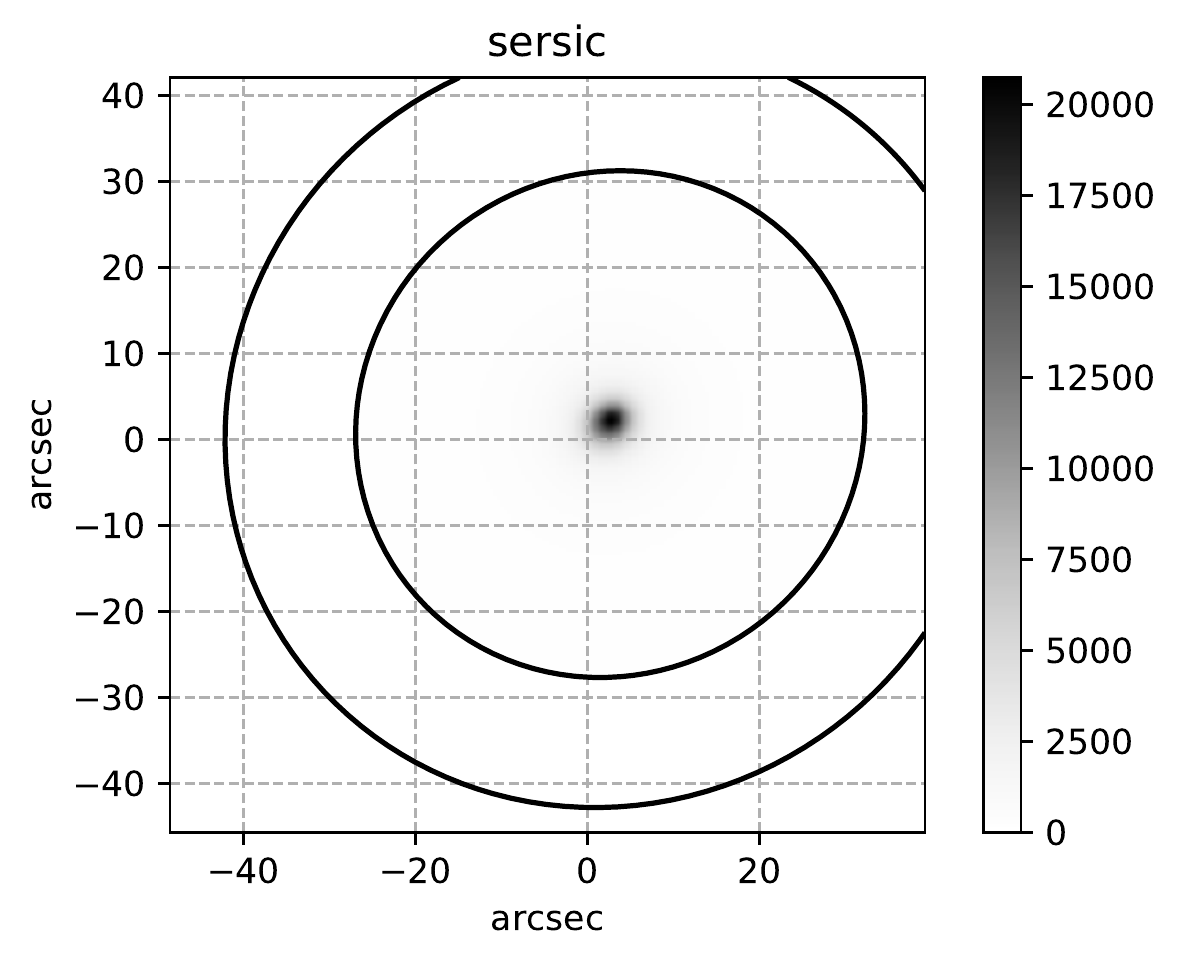} & 
\includegraphics[trim=0cm 0cm 0cm 0cm, clip=true, scale=0.45]{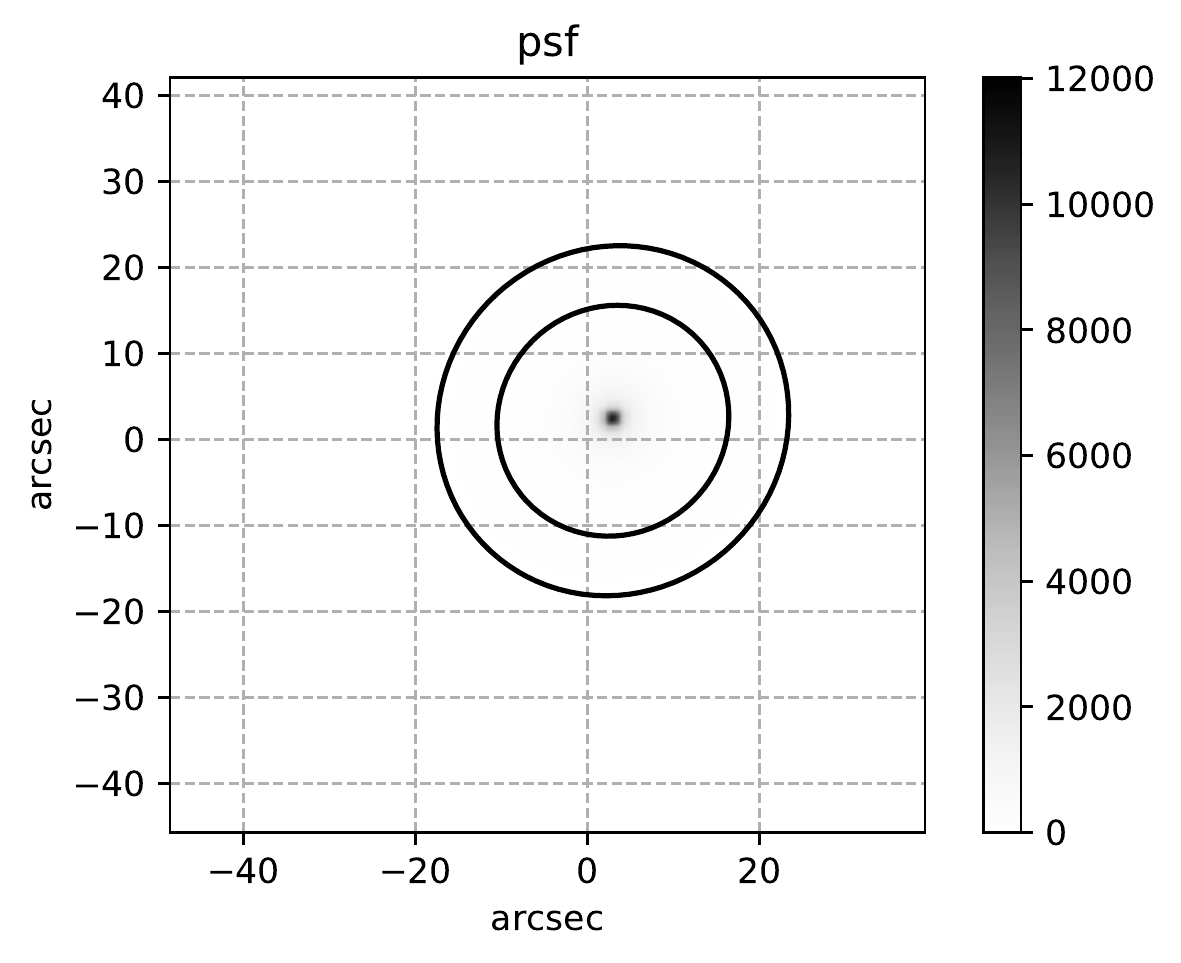}\\
\includegraphics[trim=0cm 0cm 0cm 0cm, clip=true, scale=0.45]{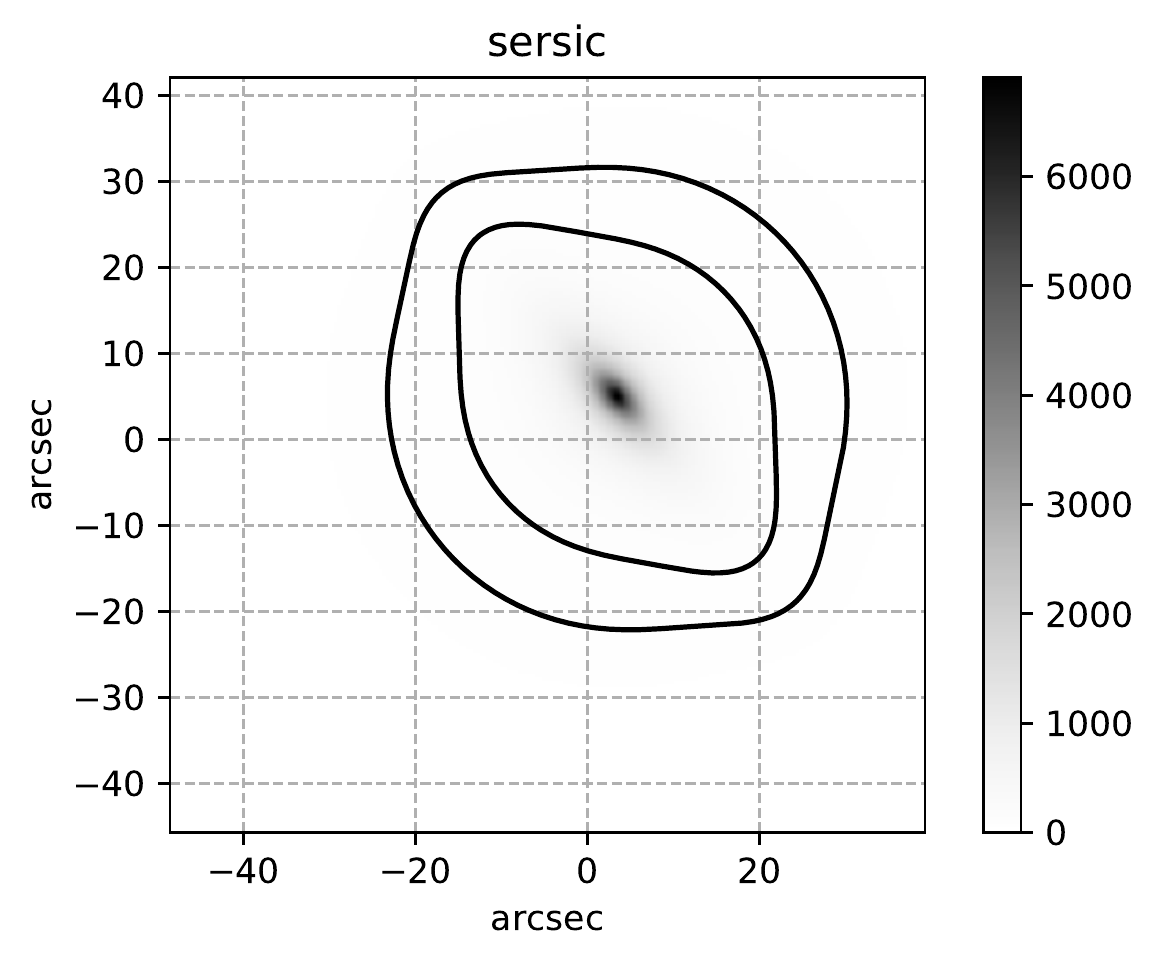} & 
\includegraphics[trim=0cm 0cm 0cm 0cm, clip=true, scale=0.45]{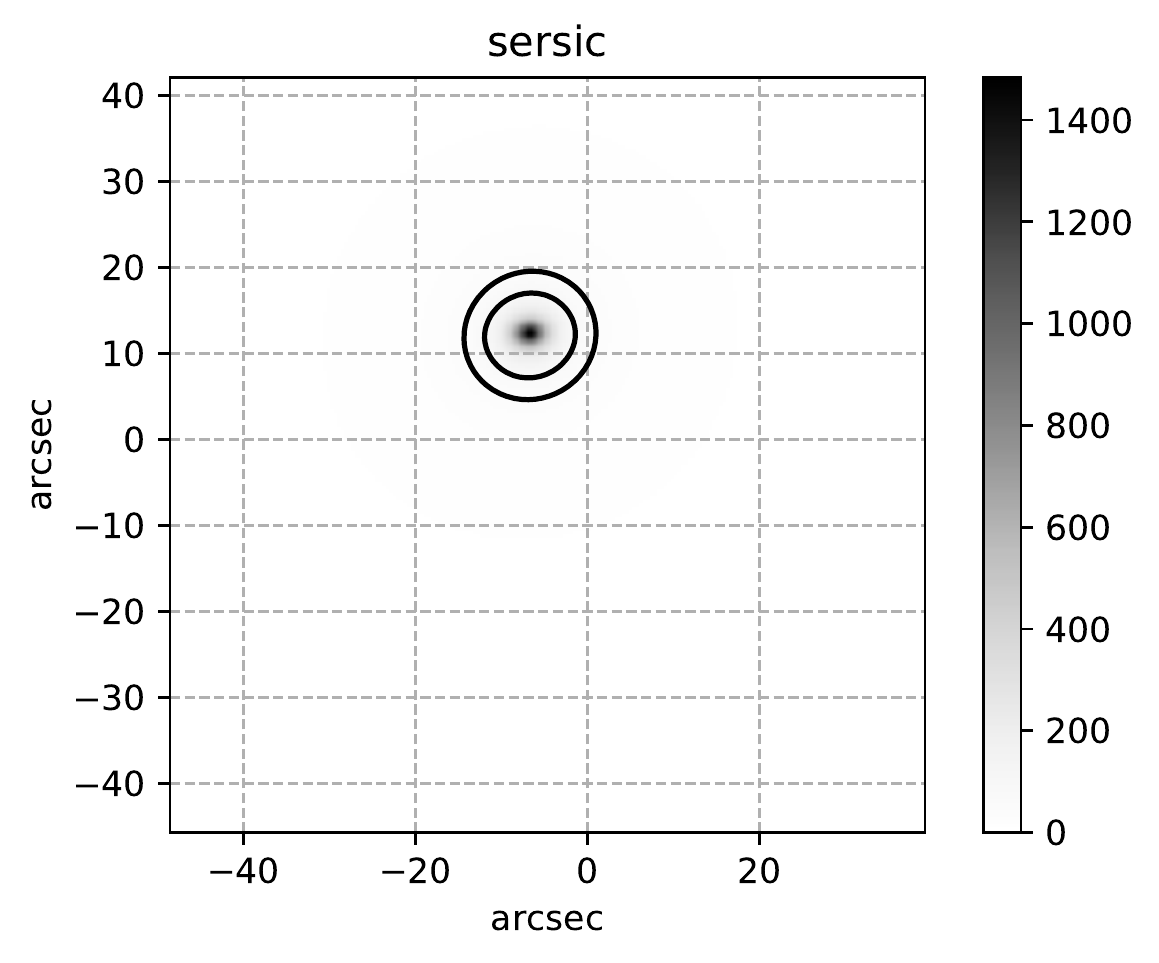} &
\includegraphics[trim=0cm 0cm 0cm 0cm, clip=true, scale=0.45]{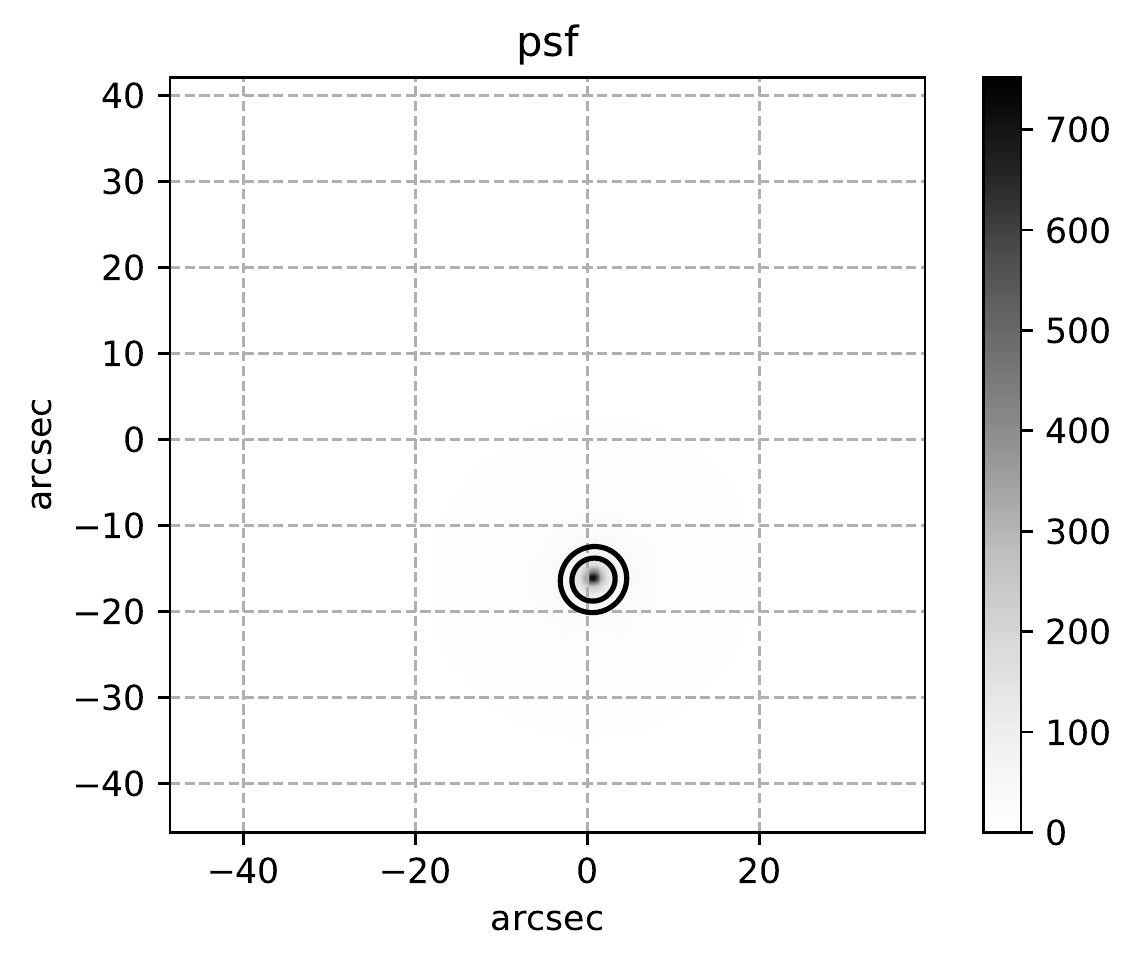} \\
\includegraphics[trim=0cm 0cm 0cm 0cm, clip=true, scale=0.45]{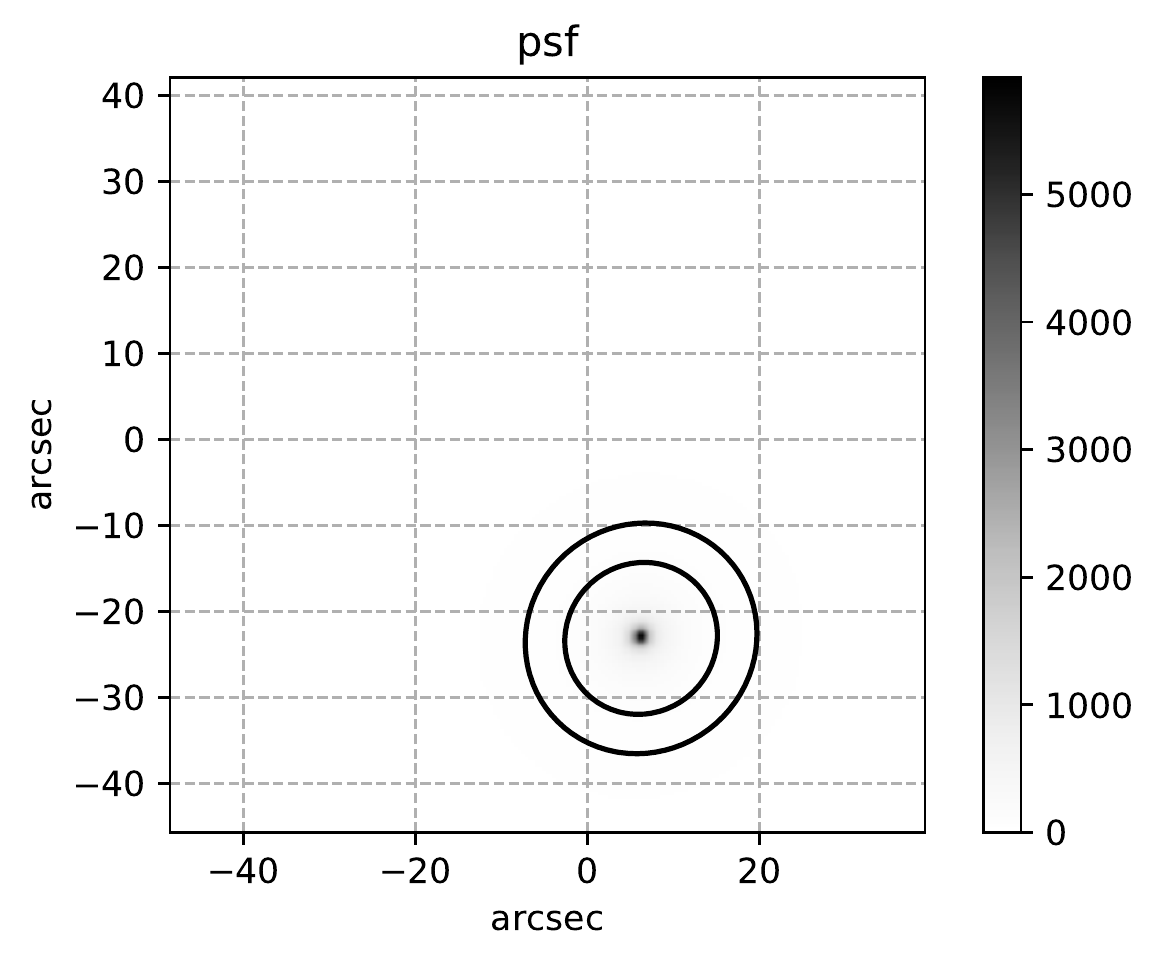} &
\includegraphics[trim=0cm 0cm 0cm 0cm, clip=true, scale=0.45]{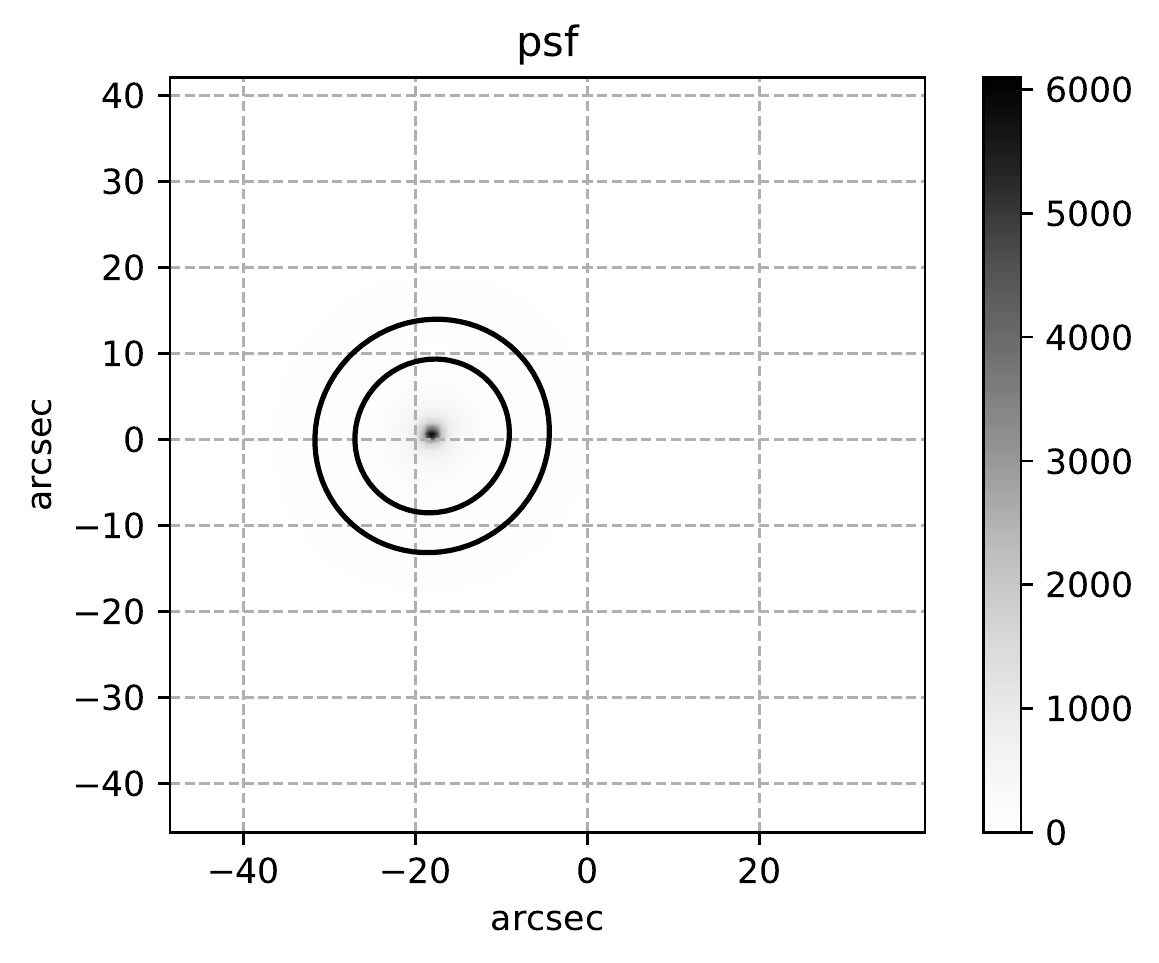}&
\includegraphics[trim=0cm 0cm 0cm 0cm, clip=true, scale=0.45]{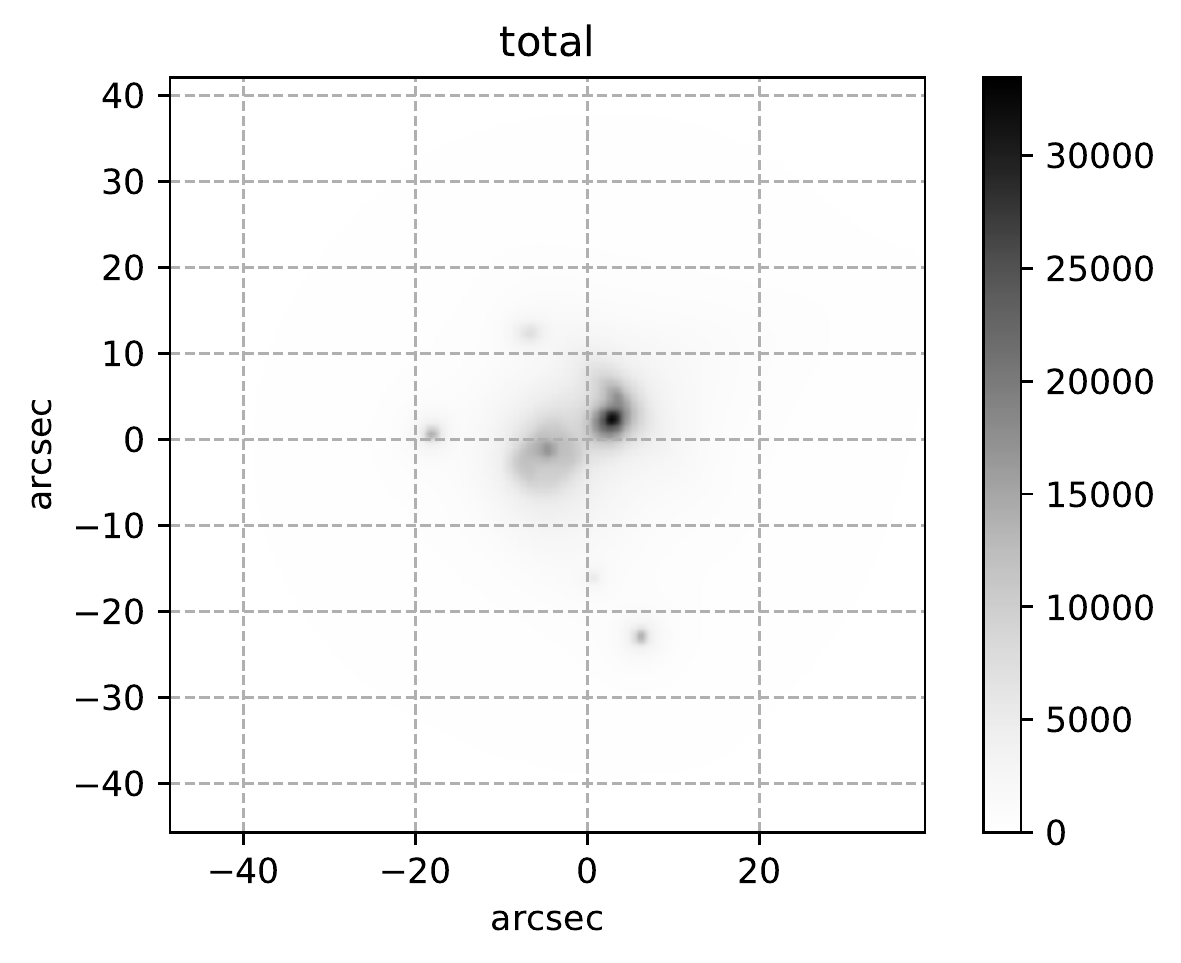}\\
\end{array}$
\caption{Components fitted via \textsc{galfit} to the i-band PanSTARRS image. For visualisation purpose, we plot the square root of the pixel values. The contours, corresponding to 40 and 80 counts, have been arbitrarily chosen to highlight the low-luminosity extension of the components. See Fig.\ref{fig: galfit_images} for a zoom-in on the central region.}
\label{fig: galfit_subc}
\end{figure*}

\begin{figure*}%[thb]
\center
\includegraphics[trim=0cm 0cm 0cm 0.1cm, clip=true, scale=0.6]{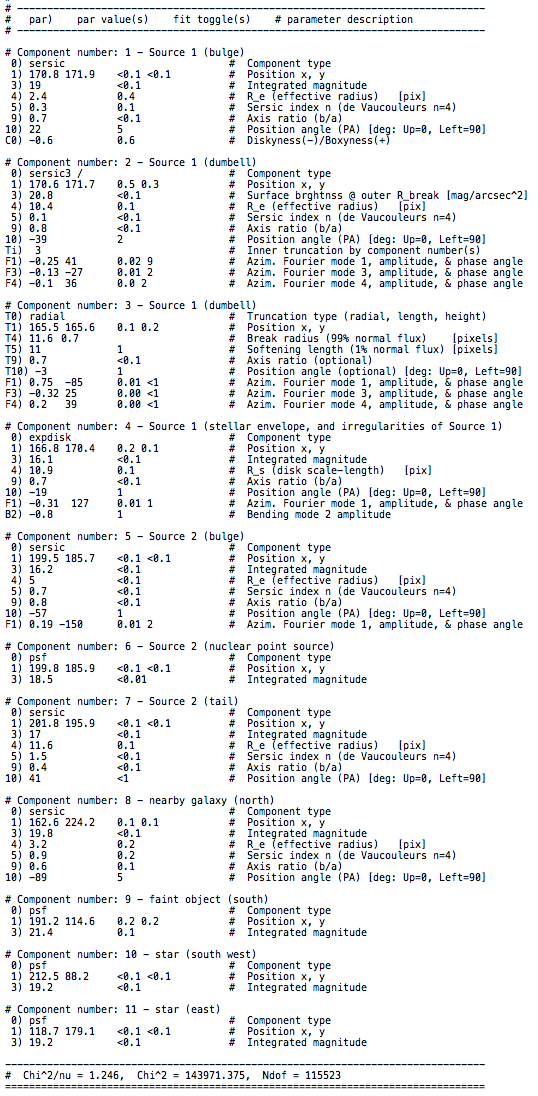}
\caption{Best fitting \textsc{galfit} parameters. Formal uncertainties (or upper limits) are indicated alongside the parameters, none of which was held fixed.}
\label{fig: galfitp}
\end{figure*}

Best fit subcomponents are plotted in Fig.\ref{fig: galfit_subc}; parameters are shown in Fig.\ref{fig: galfitp}.
% ===========
% Galfit results
% ===========

% ===========
% galfit parameters
% ===========

%%%%%%%%%%%%%%%%%%%%%%%%%%%%%%%%%%%%%%%%%%%%%%%%%%

% Don't change these lines
\bsp	% typesetting comment
\label{lastpage}
\end{document}